\documentclass[a4paper,fleqn,usenatbib,usenames,dvipsnames]{mnras} 
\usepackage{newtxtext,newtxmath} % not working for arXiv
\usepackage[T1]{fontenc}
\usepackage{ae,aecompl}
\usepackage{graphicx}	% Including figure files
\usepackage{amsmath}	% Advanced maths commands
\usepackage{comment}
\usepackage{xspace} % for \redmapper
\usepackage[dvipsnames]{xcolor} % more textcolor
\usepackage{mathtools,tabu}

\usepackage{float} % figure position
\graphicspath{{plots/}}

\newcommand{\avg}[1]{\left\langle{#1}\right\rangle}
\newcommand{\beq}{\begin{equation}}
\newcommand{\eeq}{\end{equation}}
\newcommand{\beqa}{\begin{equation}\begin{aligned}}
\newcommand{\eeqa}{\end{aligned}\end{equation}}
\newcommand{\bit}{\begin{itemize}}
\newcommand{\eit}{\end{itemize}}

\newcommand{\hiMpc}{h^{-1} \rm Mpc}

\newcommand{\hiMsun}{h^{-1} M_\odot}

\newcommand{\DS}{\Delta\Sigma}

\newcommand{\rp}{r_{\rm p}}

\newcommand{\cvir}{c_\mathrm{vir}}

\newcommand{\chisqc}{\chi^2_{\rm color}}

\newcommand{\lnM}{{\rm lnM}}
\newcommand{\lnL}{{\rm ln\lambda}}
\newcommand{\lnS}{{\rm ln\Sigma}}

\newcommand{\sigL}{\sigma_{\rm ln\lambda}}
\newcommand{\sigS}{\sigma_{\rm ln\Sigma}}

\newcommand{\Mvir}{M_{\rm vir}}
\newcommand{\Rvir}{R_{\rm vir}}
\newcommand{\cosi}{\cos(i)}
\newcommand{\lammin}{\lambda_{\rm min}}
\newcommand{\lammax}{\lambda_{\rm max}}

\newcommand{\Rlambda}{R_{\lambda}}

\newcommand{\redmapper}{redMaPPer\xspace} % \xspace add space after the command

\defcitealias{Sunayama20}{S20}
\defcitealias{DESY1CL}{Y1CL}

\title[Cluster lensing selection bias]{Optical selection bias and projection effects in stacked galaxy cluster weak lensing}
\author[DES Collaboration]{
\parbox{\textwidth}{
\Large
Hao-Yi~Wu,$^{1}$\thanks{Email: hywu@boisestate.edu}
Matteo~Costanzi,$^{2,3,4}$
Chun-Hao~To,$^{5,6,7}$
Andr\'{e}s~N.~Salcedo,$^{8}$
David~H.~Weinberg,$^{5,6,7}$ \\
James~Annis,$^{9}$
Sebastian~Bocquet,$^{10}$
Maria~Elidaiana~da~Silva~Pereira,$^{11}$
Joseph~DeRose,$^{12}$
Johnny~Esteves,$^{13}$ \\
Arya~Farahi,$^{14}$
Sebastian~Grandis,$^{10}$
Eduardo~Rozo,$^{8}$
Eli~S.~Rykoff,$^{15,16}$
Tam\'{a}s~N.~Varga,$^{17,18}$ \\
Risa~H.~Wechsler,$^{15,16,19}$
Chenxiao~Zeng,$^{5,6}$
Yuanyuan~Zhang,$^{9}$
and Zhuowen~Zhang$^{20}$
\begin{center} (DES Collaboration) \end{center}
}
\vspace{0.4cm}
\\
\parbox{\textwidth}{
$^{1}$ Department of Physics, Boise State University, Boise, ID 83725, USA\\
$^{2}$ Astronomy Unit, Department of Physics, University of Trieste, via Tiepolo 11, I-34131 Trieste, Italy\\
$^{3}$ INAF-Osservatorio Astronomico di Trieste, via G. B. Tiepolo 11, I-34143 Trieste, Italy\\
$^{4}$ Institute for Fundamental Physics of the Universe, Via Beirut 2, 34014 Trieste, Italy\\
$^{5}$ Center for Cosmology and Astro-Particle Physics, The Ohio State University, Columbus, OH 43210, USA\\
$^{6}$ Department of Physics, The Ohio State University, Columbus, OH 43210, USA\\
$^{7}$ Department of Astronomy, The Ohio State University, Columbus, OH 43210, USA\\
$^{8}$ Department of Physics, University of Arizona, Tucson, AZ 85721, USA\\
$^{9}$ Fermi National Accelerator Laboratory, P. O. Box 500, Batavia, IL 60510, USA\\
$^{10}$ University Observatory, Faculty of Physics, Ludwig-Maximilians-Universit\"at, Scheinerstr. 1, 81679 Munich, Germany\\
$^{11}$ Hamburger Sternwarte, Universit\"{a}t Hamburg, Gojenbergsweg 112, 21029 Hamburg, Germany\\
$^{12}$ Lawrence Berkeley National Laboratory, 1 Cyclotron Road, Berkeley, CA 94720, USA\\
$^{13}$ Department of Physics, University of Michigan, Ann Arbor, MI 48109, USA\\
$^{14}$ Departments of Statistics and Data Science, University of Texas at Austin, Austin, TX 78757, USA\\
$^{15}$ Kavli Institute for Particle Astrophysics \& Cosmology, P. O. Box 2450, Stanford University, Stanford, CA 94305, USA\\
$^{16}$ SLAC National Accelerator Laboratory, Menlo Park, CA 94025, USA\\
$^{17}$ Max Planck Institute for Extraterrestrial Physics, Giessenbachstrasse, 85748 Garching, Germany\\
$^{18}$ Universit\"ats-Sternwarte, Fakult\"at f\"ur Physik, Ludwig-Maximilians Universit\"at M\"unchen, Scheinerstr. 1, 81679 M\"unchen, Germany\\
$^{19}$ Department of Physics, Stanford University, 382 Via Pueblo Mall, Stanford, CA 94305, USA\\
$^{20}$ Kavli Institute for Cosmological Physics, University of Chicago, Chicago, IL 60637, USA\\
}
}

\pubyear{2022}
\begin{document}
\label{firstpage}
\pagerange{\pageref{firstpage}--\pageref{lastpage}}
\maketitle

\begin{abstract}
Cosmological constraints from current and upcoming galaxy cluster surveys are limited by the accuracy of cluster mass calibration.  In particular, optically identified galaxy clusters are prone to selection effects that can bias the weak lensing mass calibration. We investigate the selection bias of the stacked cluster lensing signal associated with optically selected clusters, using clusters identified by the \redmapper algorithm in the Buzzard simulations as a case study.  We find that at a given cluster halo mass, the residuals of \redmapper richness and weak lensing signal are positively correlated.  As a result, for a given richness selection, the stacked lensing signal is biased high compared with what we would expect from the underlying halo mass probability distribution.  The cluster lensing selection bias can thus lead to overestimated mean cluster mass and biased cosmology results.  We show that the lensing selection bias exhibits a strong scale-dependence and is approximately 20 -- 60\% for $\DS$ at large scales. This selection bias largely originates from spurious member galaxies within $\pm 20$ -- 60~$\hiMpc$ along the line of sight, highlighting the importance of quantifying projection effects associated with the broad redshift distribution of member galaxies in photometric cluster surveys.  While our results qualitatively agree with those in the literature, accurate quantitative modelling of the selection bias is needed to achieve the goals of cluster lensing cosmology and will require synthetic catalogues covering a wide range of galaxy--halo connection models.
\end{abstract}


%%%%%%%%%%%%%%%%%%%%
%%%%%%%%%%%%%%%%%%%%
\clearpage
\section{Introduction}

The number density of galaxy clusters as a function of mass and redshift is a sensitive probe for the growth rate of large scale structure and the nature of cosmic acceleration (e.g.~\citealt{Vikhlinin09, Mantz10, Rozo10, Mantz14, Bocquet15, Planck15Cluster, deHaan16, Bocquet19, Costanzi19SDSS, Costanzi21, To21b}; also see e.g.~\citealt{Frieman08, Allen11, Weinberg13, Huterer15} for reviews).  Among different observational techniques, wide-field optical imaging surveys simultaneously provide large cluster samples and allow for weak gravitational lensing mass calibration. The upcoming Vera C.~Rubin Observatory Legacy Survey of Space and Time (LSST),  Euclid, and the Nancy Grace Roman Space Telescope's High Latitude Survey have the potential to achieve unprecedented precision in cluster lensing and constraints on cosmological parameters \citep[e.g.][]{Sartoris16, Eifler21HLS, Wu21}.

Precision cosmology from optical clusters relies on unbiased mass calibration for the underlying dark matter haloes. From survey data, one first identifies clusters as overdensities of galaxies in the sky and assigns each cluster a mass proxy, e.g.~richness ($\lambda$), the number of cluster member galaxies.  Since the weak lensing signal for a single cluster usually has an insufficient signal-to-noise ratio for mass calibration, it is common to combine the lensing signal for clusters in a richness range and use this stacked signal to calibrate their mean mass \citep[see e.g.][]{Johnston07, Rozo10, Umetsu14, Simet17, Murata18, Murata19, McClintock19}.  The number counts of galaxy clusters and their mean mass are used to constrain cosmology.  In this process, biased mass calibration would lead to biased cosmological constraints.

This work focuses on the biased cluster lensing signal associated with optical selection.   In the weak lensing mass calibration process, one usually assumes that the richness-selected cluster sample has an unbiased weak lensing signal for clusters of the same mass.  However, if the richness selection preferentially includes clusters with higher lensing signals at a given mass, the stacked lensing signal and the derived mean mass would be biased high. We refer to this as the optical selection bias in cluster lensing and mass calibration.

The recent cosmological results from the Dark Energy Survey Year 1  (DES Y1) cluster abundance and lensing 
\citep[][Y1CL thereafter]{DESY1CL} suggest the presence of significant systematic bias in mass calibration  associated with the optical cluster sample defined by the \redmapper algorithm  \citep[][]{Rykoff14, Rykoff16}.  
In particular, \citetalias{DESY1CL} include an analysis that fixes the cosmology to the DES Y1 3$\times$2pt results (derived from galaxy clustering, galaxy-galaxy lensing, and cosmic shear, see \citealt{DESY1KP}) to derive the mass--observable relation using number counts.  The mass--observable relation derived this way indicates that, without correcting for the optical selection bias, the lensing mass calibration is biased high ${\sim}10\%$ for high-richness clusters ($\lambda > 30$).  On the other hand, the low-richness clusters ($\lambda < 30$) have unexpectedly low lensing signals in \citetalias{DESY1CL}.  In addition, \cite{To21b} combine the DES Y1 cluster data (abundance, lensing, and clustering) with galaxy lensing and clustering and show that the impact of selection bias on the cluster lensing signal is approximately 15\% and is scale-independent at $\gtrsim 8\ \hiMpc$.

In this work, we show that the cluster lensing selection bias largely originates from projection effects, and to a smaller extent from the effect of halo  concentration.  We use the term `projection effect' to collectively refer to changes in cluster lensing and richness due to matter and galaxies projected along the line of sight.  The impact of projection effects on lensing can be further split into the non-spherical matter distribution inside the halo's virial radius and the matter in filaments in large scale structure. We will sometimes refer to the former as orientation bias, but we consider it to be a subset of projection effects, and the division between these two contributions is only approximate.  Simulations have shown that dark matter haloes tend to have triaxial shapes, i.e.~3D ellipsoids described by three different axis lengths \citep[see e.g.][]{JingSuto02, KasunEvrard05, Bett12, Wu13, Bonamigo15}. If a cluster sample is dominated by clusters with major axes aligned along the line of sight, then the stacked lensing signal would be boosted relative to the prediction based on spherically symmetric haloes \citep[e.g.][]{Gavazzi05, Oguri05, CorlessKing07, Limousin13, Dietrich14, Osato18, Herbonnet21, Park21, ZZhang22}.

Projection effects change cluster richness by including galaxies outside the halo virial radius in the richness calculation \citep[e.g.][]{Erickson11, Kohlinger15, Rozo15RM4, Farahi16, Wojtak18, Costanzi19projection, Myles20projection}.  In imaging surveys, cluster members are identified by their colours, and galaxies matching the colour criterion but outside the cluster virial radius can be misidentified as members.  These galaxies are associated with uncorrelated background or correlated large scale structure, and the latter tends to increase the lensing signal and richness simultaneously.

The cluster lensing selection bias can be mathematically described as the correlation between richness residual and lensing residual at a fixed mass; that is, after removing the mass dependence of richness and lensing, a cluster with a large positive richness residual tends to also have a large positive lensing residual.  If our cluster sample includes all possible richness values at a given mass, then we expect the stacked lensing signal to be unbiased.  However, it is necessary to select clusters above a certain richness limit because clusters with very low richness tend to be spurious.  Therefore, in the presence of correlated residuals, a richness selection would lead to a biased stacked lensing signal.

Given that projection effects coherently impact both richness and lensing signal, it is imperative to study richness and lensing simultaneously using simulations that self-consistently model cluster richness and lensing signal.  In this work, we use the Buzzard simulations for DES \citep{DeRose19, DeRose22}, with cluster samples defined by the \redmapper cluster finding algorithm.  We first mimic the cluster selection in \citetalias{DESY1CL} and quantify the lensing selection bias. We then investigate the root cause of the selection bias by examining the underlying dark matter and galaxy distributions.

Recently, \citet[][S20 thereafter]{Sunayama20} use a halo occupation distribution (HOD) model to populate galaxies in N-body simulations and show that a \redmapper-like cylindrical member selection leads to a 20\% lensing bias for scales $\gtrsim 10\ \hiMpc$.  As we will discuss in detail in Section~\ref{sec:discussion}, the lensing selection bias they find is in general lower than ours, and we attribute this to the differences in the galaxy models.

This paper is organised as follows.  We briefly review cluster weak lensing and introduce the selection bias formalism in Section~\ref{sec:basics} and describe our simulations in Section~\ref{sec:sims}.  Section~\ref{sec:results} presents our main results on the cluster lensing selection bias.  We investigate the correlation between richness residual and lensing residual in Section~\ref{sec:correlation} and the link between projection effects and selection bias in Section~\ref{sec:cylinder}.  Our results are discussed in Section~\ref{sec:discussion} and summarised in Section~\ref{sec:summary}.

We put most technical details and robustness tests in extensive appendices. Appendix~\ref{app:buzzard_versions} compares different versions of the Buzzard simulations, and Appendix~\ref{app:diagnosis} compares different diagnoses for calculating the selection bias.  Appendix~\ref{app:alternatives} shows that our results are robust against member galaxy selection criteria. Appendices~\ref{app:triaxiality} and~\ref{app:concentration} address the impact of halo triaxiality and concentration.

Throughout the paper we use the flat $\Lambda$CDM cosmology implemented in the Buzzard simulations:  $\Omega_M$ = 0.286, $h$ = 0.7, $\sigma_8$ = 0.82, $n_s$ = 0.96, $\Omega_B$ = 0.046, and $N_{\rm eff}$ = 3.046. All projected distances are in {\em physical} Mpc without $h$ (denoted as pMpc)\footnote{Using physical instead of comoving distances is somewhat unusual for a simulation-based cluster study.  We have made this choice because the \redmapper cluster boundaries are defined in physical units.  We have checked that for our redshift bins of $\Delta z = 0.15$, stacking clusters using physical distances and comoving distances leads to negligible differences.}, while all line-of-sight distances are in comoving $\hiMpc$.  All halo masses are virial mass $\Mvir$ (in the unit of $\hiMsun$) based on the redshift-dependent spherical overdensity provided in \cite{BryanNorman98}.

%%%%%%%%%%%%%%%%%%%%
%%%%%%%%%%%%%%%%%%%%
\section{Formalism of cluster counts, weak lensing, and selection bias}
\label{sec:basics}

In this section, we briefly describe the formalism for modelling cluster number counts and stacked weak lensing, and we extend the formalism to model the selection bias.  For comprehensive reviews for gravitational lensing, we refer readers to \cite{BartelmannSchneider01, Kilbinger15, Umetsu20}.

The weak lensing signal of a galaxy cluster is related to its excess surface mass density $\DS$, the surface density contrast at a projected distance $\rp$,
\beq
\DS (\rp) = \bar{\Sigma}(< \rp) - \Sigma(\rp)  \ , 
\label{eq:DS}
\eeq
where 
$\bar{\Sigma}(<\rp)$ is the cumulative mean surface mass density {\em within} $\rp$, and
$\Sigma(\rp)$ is the differential mean surface mass density at $\rp$; $\Sigma(\rp)$ can be calculated by integrating the line-of-sight 3D mass density distribution $\rho(r)$. Below we show the expressions for $\Sigma$ and note that the expressions for $\DS$ can be derived analogously.

We focus on the number density and the mean lensing signal for a cluster sample defined by the richness range ($\lammin$, $\lammax$) at a given redshift.  The comoving number density of this sample is given by
\beq
n(\lammin, \lammax) = \int_{\lammin}^{\lammax}d\lambda \int dM \frac{dn}{dM} P(\lambda|M)  \ ,
\eeq
where $P(\lambda | M )$ denotes the probability distribution function (PDF) of $\lambda$ at a given mass, and $dn/dM$ denotes the halo mass function and is determined by cosmological parameters. Due to the strong degeneracy between the parameters determining $dn/dM$ and $P(\lambda|M)$, any bias in the latter will lead to biased cosmological parameters.

To calibrate $P(\lambda | M)$, we use the stacked lensing signal of the same cluster sample, which is given by  
\beq
\Sigma(\lammin, \lammax) = \frac
{\int_{\lammin}^{\lammax}d\lambda \int dM \frac{dn}{dM} P(\lambda|M) \avg{\Sigma|M} }
{\int_{\lammin}^{\lammax}d\lambda \int dM \frac{dn}{dM} P(\lambda|M)}  \ ,
\eeq
where $\avg{\Sigma | M }$ is the mean lensing signal at a given mass, 
\beq
\avg{\Sigma | M } = \int d\Sigma\ \Sigma\ P(\Sigma|M)  \ . 
\eeq
The equations above assume that $\lambda$ and $\Sigma$ are uncorrelated.  However, if $\Sigma$ is correlated with $\lambda$ at a given mass, $P(\lambda | M)$ and $P(\Sigma | M)$ need to be replaced by a joint probability distribution  $P(\lambda, \Sigma | M)$.  The stacked lensing signal is then given by
\beq
\Sigma(\lammin, \lammax) \propto
\int_{\lammin}^{\lammax}d\lambda \int dM \frac{dn}{dM} 
\int d\Sigma\ \Sigma\ P(\lambda, \Sigma|M)   \ .
\eeq

To proceed, we need to model $P(\lambda, \Sigma|M)$.  It has been shown that at fixed mass, $\lambda$ is well described by a log-normal distribution \citep[e.g.][]{Anbajagane20, To21b}. Similarly, we have verified from our simulations that $\Sigma$ at fixed mass can be described by a log-normal distribution. To account for the correlation between the two observables, we consider a bivariate log-normal distribution. We note that deviations from this assumption could exist and will require further numerical modelling.

Below we analytically model the selection bias by assuming that $\lnL$ and $\lnS$ follow a bivariate Gaussian distribution.  Let us assume this joint PDF has a mean
$\bigg(\avg{\lnL \big| M}, \avg{\lnS \big| M}\bigg)$ and a covariance matrix 
\beq
\mathcal{C} = 
 \begin{pmatrix}
 \sigL^2       &  r\sigL\sigS \\
 r\sigL\sigS   &  \sigS^2
 \end{pmatrix}  
\ ,
\eeq  
where $\sigL$ and $\sigS$ are the standard deviations of the two observables at a given mass, and $r$ is the correlation coefficient between the residuals, ${\rm corr}[\lnL - \avg{\lnL | M}, \lnS - \avg{\lnS | M}]$.
The PDF of the lensing signal associated with the richness selection is given by
\beqa
& p\bigg(\lnS \big| \lnL , M \bigg) 
= \frac{P\bigg(\lnS, \lnL \big| M \bigg)}
       {P\bigg(\lnL       \big| M \bigg)} \\
& \propto \frac
 {\mbox{bivariate Gaussian}\bigg(\big(\avg{\lnL \big| M}, \avg{\lnS \big| M}\big) , \mathcal{C} \bigg) } 
 {\mbox{Gaussian} \bigg(\avg{\lnL \big| M} , \sigL^2 \bigg) }  \ .
\eeqa
Under this assumption, the conditional probability distribution of $\lnS$ given $\lnL$ corresponds to a Gaussian distribution with the mean
\beq
\avg{\lnS \big| \lnL,  M } = 
\avg{\lnS \big| M} 
+ r \ \sigS \ 
\frac{\big(\lnL - \avg{\lnL \big| M}\big)}{\sigL}
  \ .
\label{eq:bias_corr}
\eeq
The last term describes the selection bias associated with cluster lensing (also see \citealt{WhiteM10, Evrard14} for similar derivations). This formalism is also mathematically equivalent to the impact of halo assembly bias on cluster lensing and clustering \citep[see e.g.][]{Wu08}.

The correlation between the $\lambda$ residual and the $\Sigma$ residual at a fixed mass is the essence of the cluster weak lensing selection bias.  In Section~\ref{sec:correlation} we will quantify the correlated residuals in the mock \redmapper cluster sample.

%%%%%%%%%%%%%%%%%%%%
%%%%%%%%%%%%%%%%%%%%
\section{Galaxy clusters in the Buzzard Simulations}\label{sec:sims}

In this section we describe our simulated \redmapper clusters from the Buzzard simulations.

%%%%%%%%%%%%%%%%%%%%
\subsection{The Buzzard simulations}

The Buzzard simulations \citep{DeRose19, DeRose22} include a suite of synthetic catalogues based on the {\sc addgals} algorithm \citep{Wechsler21} and are designed for supporting the DES data analyses. Below we describe the simulation framework, and in Appendix~\ref{app:buzzard_versions} we describe the particular versions of Buzzard we use.

The first step of the simulation creates a galaxy catalogue with r-band magnitudes. The algorithm performs subhalo abundance matching between an N-body simulation with well-resolved subhaloes and the observed luminosity function in the r-band.  Using the resulting galaxy catalogue, the algorithm calibrates (1) the relation between mass and r-band magnitude for central galaxies, and  (2) the relation between local density and r-band magnitude for satellite galaxies. These relations are then used to assign galaxies to resolved haloes or dark matter particles in a large light-cone simulation with a lower resolution.

The second step of the simulation assigns colour to each galaxy.  At a given r-band magnitude, galaxies from the observed catalogue are ranked by their g--r colour, and galaxies from the simulated catalogue are ranked by an environmental proxy, such as the distance to the nearest neighbour.  The algorithm then performs abundance matching between these two ranked lists with some scatter, and the observed galaxy spectral energy distributions (SEDs) are assigned to the corresponding galaxies in the simulated catalogue.

In this work, we use 12 realisations of the DES Y1 data (1120 deg$^2$ each) based on Buzzard version 1.9.2+1 and one realisation of the DES Y3 data (4143 deg$^2$) based on Buzzard version 1.9.8 \citep[presented in][respectively]{DeRose19,DeRose22}.  In Appendix~\ref{app:buzzard_versions} we compare the two versions.   The key difference relevant for this work is that the Y1 realisations have a narrower red sequence than the Y3 realisation. We find that the lensing selection bias results from these two versions are statistically consistent, and therefore in the main text we combine the results from both versions.

\cite{Wechsler21} show that the massive haloes in Buzzard tend to have fewer member galaxies than observed, and this deficit is attributed to the artificial disruption of subhaloes in dense environments in the N-body simulation used for subhalo abundance matching.  As a result, clusters in Buzzard tend to have lower richness values compared with observed clusters of similar mass, and above a richness threshold Buzzard has fewer clusters than observed.  In addition, \cite{To21b} show that the selection bias in Buzzard is higher than that indicated by the DES Y1 data at large scales, which can also be attributed to the low galaxy density in clusters in Buzzard.  Because of this discrepancy, we do not directly use the selection bias derived from Buzzard to correct the observed lensing signal.  Instead, we use Buzzard to study the nature of selection bias and leave to future works a full calibration using suites of galaxy--halo connection models.

%%%%%%%%%%%%%%%%%%%%
\subsection{The \redmapper cluster finding algorithm}

The \redmapper algorithm \citep{Rykoff14,Rykoff16} identifies galaxy clusters in a photometric galaxy catalogue based on the red sequence, i.e.~the tight colour--magnitude relation for galaxies in clusters.  The algorithm first calibrates a red-sequence template using a sample with both photometric and spectroscopic information.  This template is then used to select red galaxies as possible central galaxies.

For a central galaxy, the algorithm finds its candidate member galaxies and assigns each member a membership probability, the probability that a galaxy is a true cluster member calculated based on its magnitude, colour, and distance to the central galaxy.  The member galaxies are selected in a projected aperture $\Rlambda$, which is iteratively calculated to match the relation $\Rlambda =  1 (\lambda/100)^{0.2} \ \hiMpc $ (physical).  This relation has been calibrated to minimise the scatter of $L_X$ given $\lambda$ \citep{Rykoff12}.

In a process called percolation, all possible central galaxies are ranked by a preliminary richness, and a higher-ranked central galaxy is prioritised in obtaining its members; that is, for a candidate galaxy member in the vicinity of two possible central galaxies, it will be assigned a higher membership probability to the higher-ranked central.

In this work we use the \redmapper version 6.4.22, which is essentially the same version as used in DES \citetalias{DESY1CL} (see \citealt{McClintock19}).  We use the halo centres as cluster centres to calculate the richness to avoid mismatched halo--cluster pairs and misidentified central galaxies \citep[e.g.][]{Zhang19}.

%%%%%%%%%%%%%%%%%%%%
\subsection{Measuring stacked cluster lensing in simulations}
%%%%%%%%%%%%%%%%%%%%
\begin{figure}
\centering
\includegraphics[width=\columnwidth]{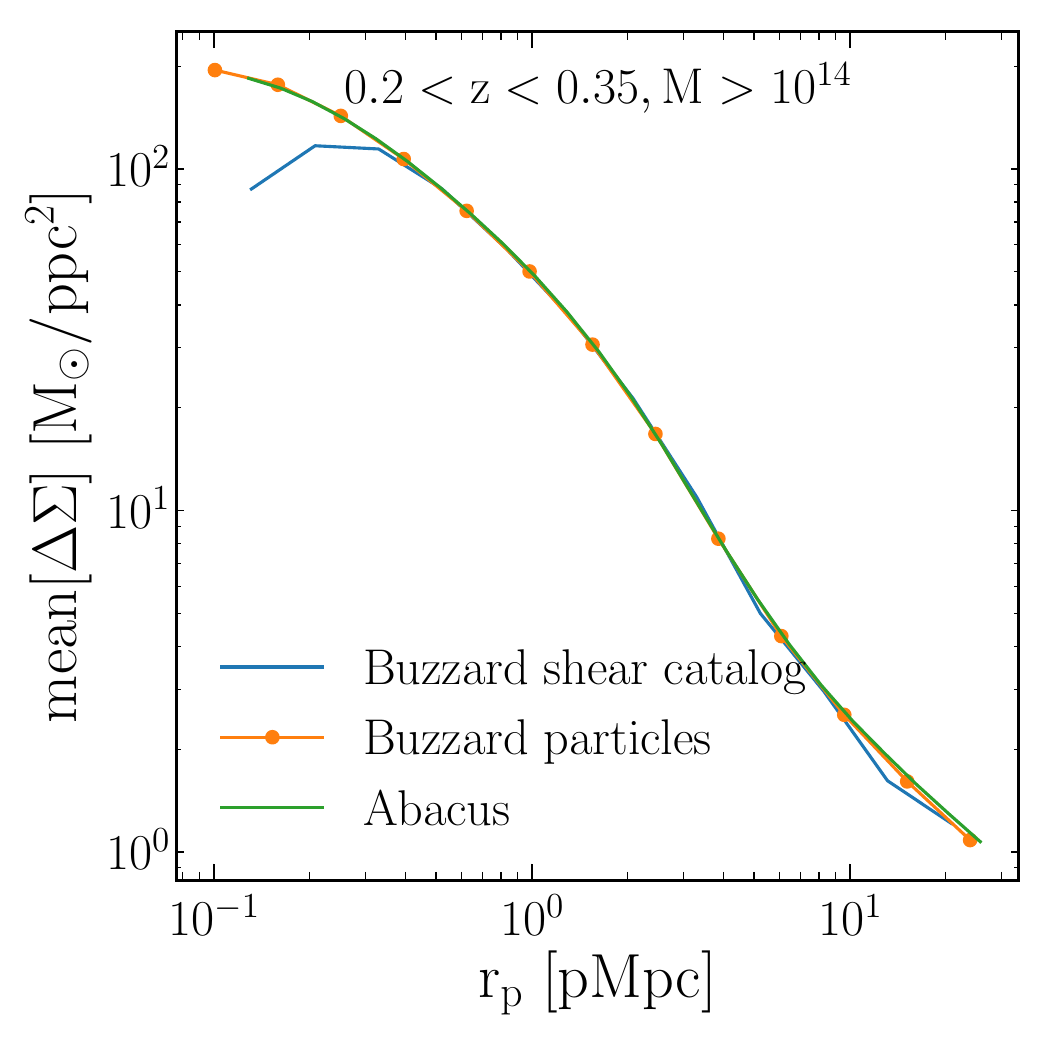}
\caption[]{Mean $\DS$ profiles from the Buzzard simulations, calculated using dark matter particles and the ray-tracing shear catalogue, compared with the Abacus Cosmos simulations with 3 times higher mass resolution.  The Buzzard particle calculations have adequate resolution down to 0.1 pMpc.  The pMpc and ppc in the axis labels refer to physical megaparsec and parsec.}
\label{fig:DS_demo}
\end{figure}
%%%%%%%%%%%%%%%%%%%%

We use dark matter particles to calculate the surface mass density $\Sigma$ and excess surface density $\DS$ of clusters, using cylinders of depth comoving $\pm 100\ \hiMpc$. This projection depth is sufficient to account for the correlated structure along the line of sight (also see \citealt{Osato18}), and we have tested that using  cylinders of $\pm 200 \ \hiMpc$ or using particles in cones leads to negligible differences.  Because Buzzard joins two different boxes at $z=0.32$, we discard haloes within $\pm 100 \ \hiMpc$ of this discontinuity boundary.  We calculate the profiles for all haloes in the parent N-body simulations with $\Mvir \ge 10^{13}\hiMsun$, regardless of whether a halo is in the mock cluster catalogue or not.  As we will show below, we use all these haloes to form a control sample for lensing signals.  We additionally calculate the triaxial shape of each halo using the dark matter particles within $\Rvir$ (see Appendix~\ref{app:triaxiality}).

Fig.~\ref{fig:DS_demo} compares the $\DS$ profiles for haloes with $0.2 < z < 0.35$ and $\Mvir \ge 10^{14}\ \hiMsun$, calculated from particles and from the ray-tracing shear catalogue.  We use the true redshift and shear in the shear catalogue, ignoring photometric errors and intrinsic galaxy ellipticities.  The ray-tracing and particle calculations agree with each other at large scales, while the former does not have sufficient resolution below 0.4 pMpc.  To test the resolution limit of Buzzard we use the Abacus Cosmos simulations \citep{Garrison18}; we use the dark matter particles in the 720 $\hiMpc$ boxes at $z=0.3$, which have 3 times better mass resolution than Buzzard.\footnote{The Abacus Cosmos simulations are based on $M_{200m}$, a different cosmology, and a different redshift, and thus we only use them for resolution comparison.}  The comparison with Abacus shows that the spatial resolution of Buzzard is adequate down to 0.1 pMpc. Scales below 0.1 pMpc are not usually included in weak lensing cluster mass calibration.

%%%%%%%%%%%%%%%%%%%%
%%%%%%%%%%%%%%%%%%%%
\section{Quantifying the selection bias in the stacked \redmapper cluster lensing signal}\label{sec:results}

%%%%%%%%%%%%%%%%%%%%
\begin{figure*}
\centering
\includegraphics[width=2\columnwidth]{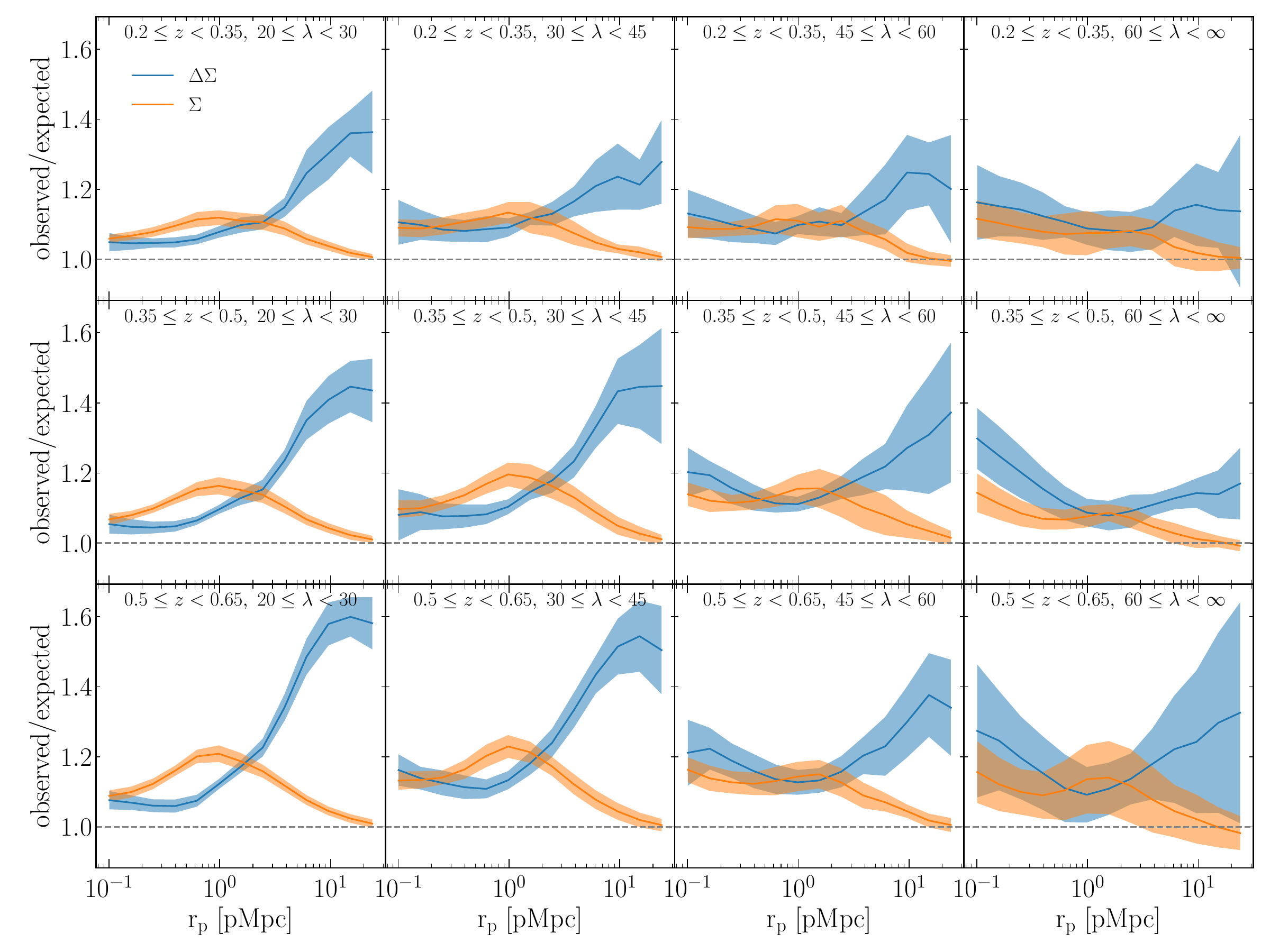}
\caption[]{Cluster lensing selection bias of $\Sigma$ and $\DS$, quantified by the ratio between the signal from a richness-selected sample (observed) and the signal expected from the underlying halo mass PDF (expected). We present different redshift bins (rows) and richness bins (columns).  For $\Sigma$, the selection bias peaks at approximately 1 pMpc and vanishes at large scales.  For $\DS$, the selection bias includes the small scale effects and can be as high as 20 -- 60\% at large scales.}
\label{fig:bias_DS}
\end{figure*}
%%%%%%%%%%%%%%%%%%%
\begin{figure}
\includegraphics[width=1\columnwidth]{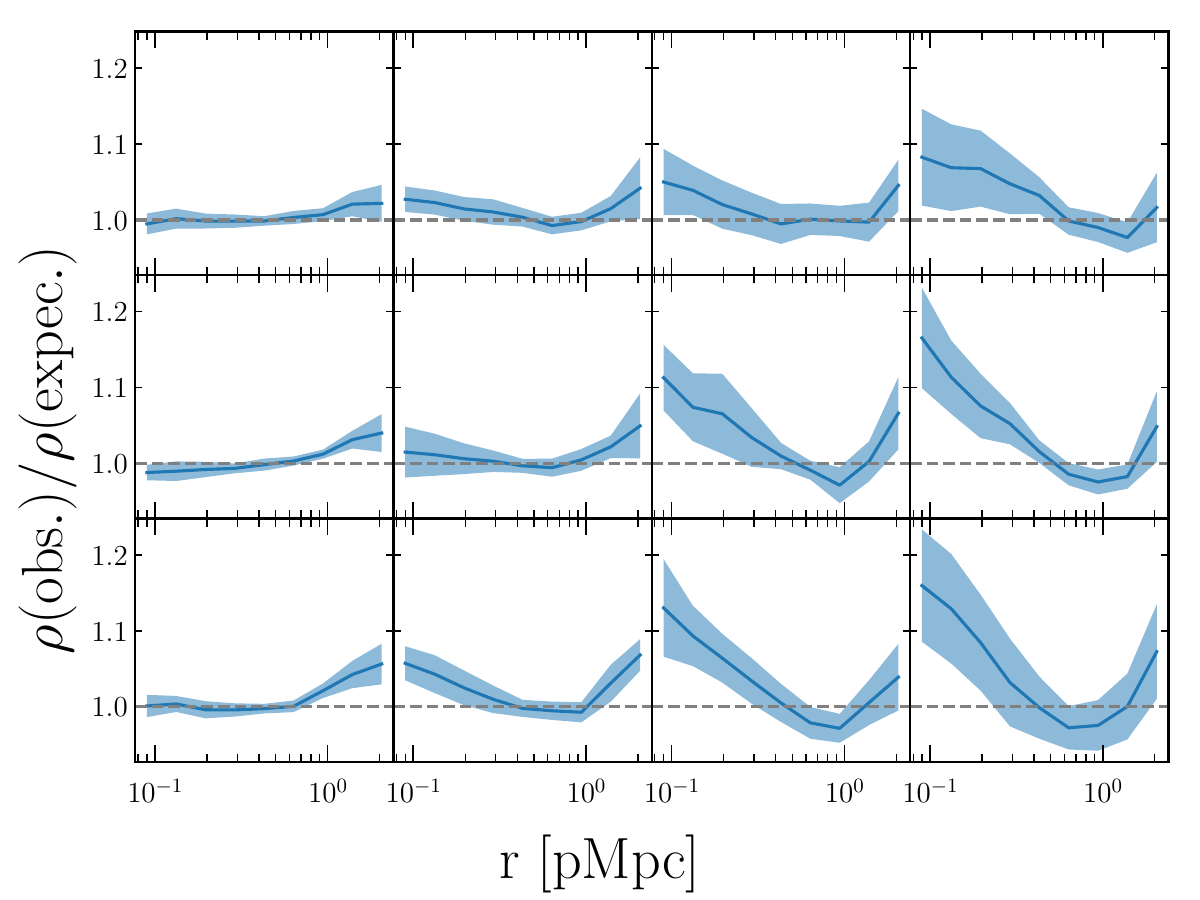}
\caption[]{Selection bias in 3D density profiles $\rho(r)$, analogous to Fig.~\ref{fig:bias_DS} with matching panels.  While low-richness clusters exhibit negligible selection bias in $\rho$, high-richness clusters show $\sim$ 10 -- 15\% selection bias in $\rho$ at small scales.  This explains the different scale-dependence of $\Sigma$ bias for low- and high-richness clusters seen in Fig.~\ref{fig:bias_DS}.
}
\label{fig:bias_rho}
\end{figure}
%%%%%%%%%%%%%%%%%%%%

With the simulated cluster lensing signal described above, we calculate the stacked lensing signal in redshift and richness bins.  We use the same binning as in DES \citetalias{DESY1CL}: three redshift bins bounded by (0.2, 0.35, 0.5, 0.65), and four richness bins bounded by (20, 30, 45, 60, $\infty$).

To quantify the cluster lensing selection bias, we compare the stacked lensing signal of clusters selected by richness with the signal expected from the underlying dark matter halo mass PDF.  This `expected' signal accounts for the scatter between richness and mass but assumes uncorrelated residuals between richness and lensing at a given mass.  In Appendix~\ref{app:diagnosis}, we detail three methods for calculating the expected lensing signal from this mass PDF and show that they give consistent results.  In the main text, we present the `weighting' method.

Fig.~\ref{fig:bias_DS} shows the selection bias for $\Sigma$ and $\DS$.  We show the mean and standard deviation calculated from the 12 realisations of the DES Y1 data and the one realisation of the DES Y3 data, weighted by the area.  For $\Sigma$, all richness bins exhibit biases of approximately 10 -- 20\%, with strong scale-dependence.  For the low-richness bins, the selection bias peaks at approximately 1 pMpc and is weak at small and large scales.  For the high-richness bins, the selection bias is substantial at small scales. In all cases, the selection bias for $\Sigma$ vanishes at scales $\gtrsim 20$ pMpc.  In contrast to $\Sigma$, the bias of $\DS$ is non-vanishing at large scales because $\DS$ at each $\rp$ contains the information of $\Sigma$ from $r < \rp$ (equation~\ref{eq:DS}). The bias in $\DS$ can be as high as 20 -- 60\% at large $\rp$.

We would like to understand to what extent the biased lensing profile presented in Fig.~\ref{fig:bias_DS} is caused by a biased 3D density profile.  To answer this, we repeat the selection bias  calculation for the 3D density profile $\rho(r)$.  For each halo, we calculate the spherically averaged $\rho(r)$ by counting dark matter particles in spherical shells around the halo centre.  We then calculate the mean $\rho$ in a richness bin and the $\rho$ expected from haloes with the same mass PDF.

Fig.~\ref{fig:bias_rho} shows the selection bias of 3D density profiles $\rho$ out to 3 pMpc.  For low-richness clusters, the selection bias of $\rho$ is negligible, while for high-richness clusters, the small scales exhibit a $\sim10\%$ selection bias in $\rho$.  This difference in small-scale behaviour explains the difference between high- and low-richness clusters shown in Fig.~\ref{fig:bias_DS}.  For low-richness clusters, the small-scale lensing selection bias is associated with the 2D projection, while for high-richness clusters, part of the small-scale  selection bias is due to the biased 3D density profiles.

Fig.~\ref{fig:bias_rho} implies that our high-richness sample preferentially selects haloes with higher 3D density at small scales at a given mass.  We expect that these haloes have higher concentrations.  We investigate the influence of halo concentration in Appendix~\ref{app:concentration}. Fig.~\ref{fig:cvir_pdf} compares the concentration distribution for a richness-selected sample and for a sample with the same mass PDF.  As expected from Fig.~\ref{fig:bias_rho}, the high-richness clusters tend to have higher concentrations than haloes of the same mass PDF, while the low-richness clusters do not show such a bias.  Fig.~\ref{fig:Sigma_con} shows how concentration affects the lensing profile.  The dependence of $\Sigma$ on halo concentration is quite different from what we see in Fig.~\ref{fig:bias_DS}, and therefore the halo concentration has limited predictive power for the selection bias.

We also examine the selection bias associated with the triaxial halo orientation in Appendix~\ref{app:triaxiality}.  Similar to concentration, our cluster sample preferentially selects haloes with major axes parallel to the line of sight.  However, the orientation and concentration selection cannot fully account for the selection bias we find (Fig.~\ref{fig:bias_Sigma_extra_prop}).

In the next sections, we will investigate the origin of the selection bias.  In Section~\ref{sec:correlation} we will calculate the correlation between the richness residual and the lensing residual at a given halo mass.  In Section~\ref{sec:cylinder}, we will examine the relation between projection effects and selection bias.

%%%%%%%%%%%%%%%%%%%%
%%%%%%%%%%%%%%%%%%%%
%%%%%%%%%%%%%%%%%%%%
\section{Correlation between the richness residual and the lensing residual at a given halo mass}\label{sec:correlation}

%%%%%%%%%%%%%%%%%%%%
\begin{figure*}
\includegraphics[width=2\columnwidth]{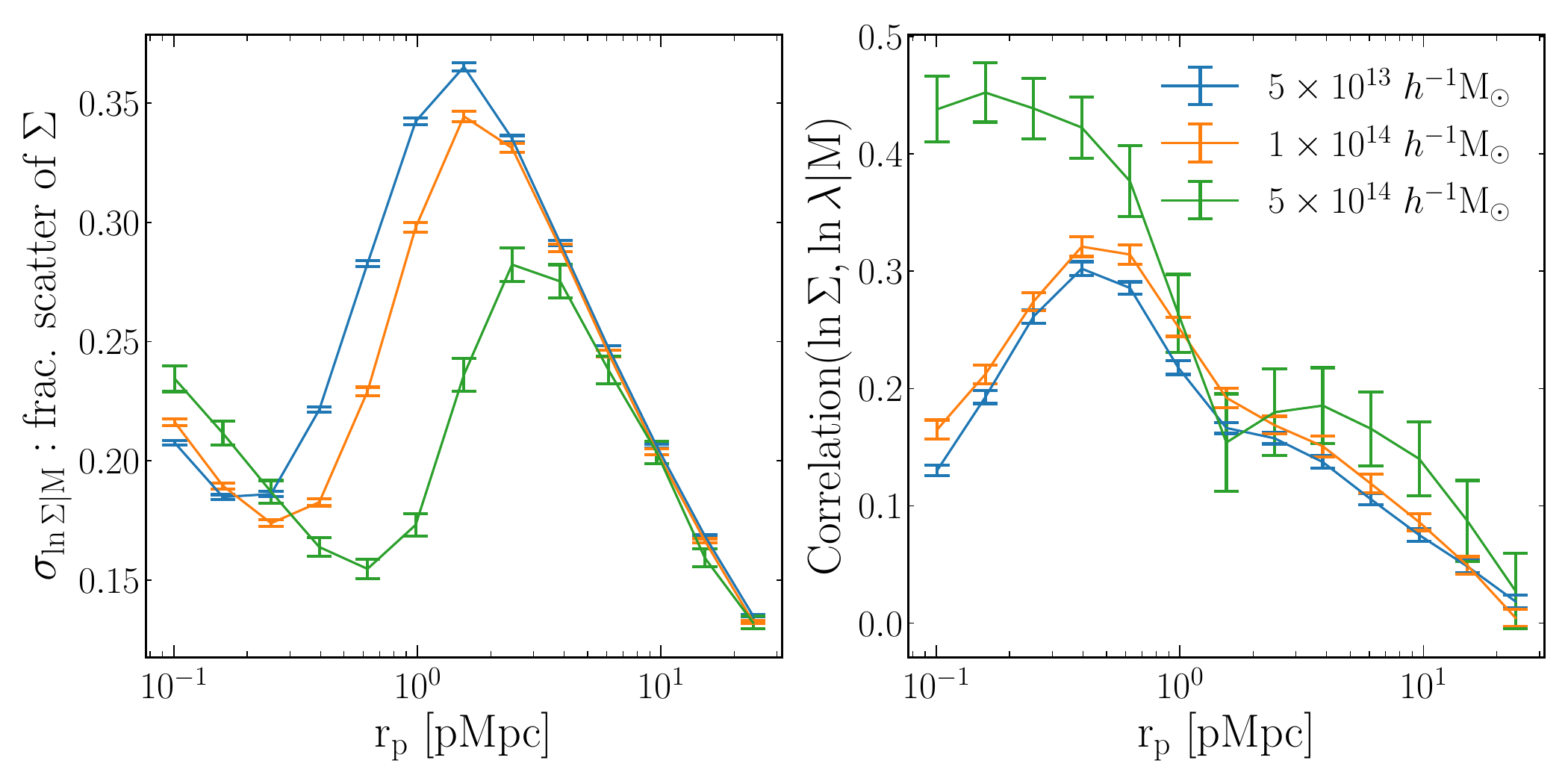}
\caption[]{Left: scatter of the lensing signal at a given mass as a function of projected radius, for $0.2 \le z < 0.35$ and 3 halo masses.   Right: correlation between the lensing residual and the richness residual at a given mass, as a function of projected radius.  The selection bias is proportional to the product of the lensing scatter and the correlation coefficient (equation~\ref{eq:bias_corr}), and their scale-dependence explains the scale-dependence of the selection bias seen in Fig.~\ref{fig:bias_DS}.
}
\label{fig:corr}
\end{figure*}
%%%%%%%%%%%%%%%%%%%%
\begin{figure}
\includegraphics[width=\columnwidth]{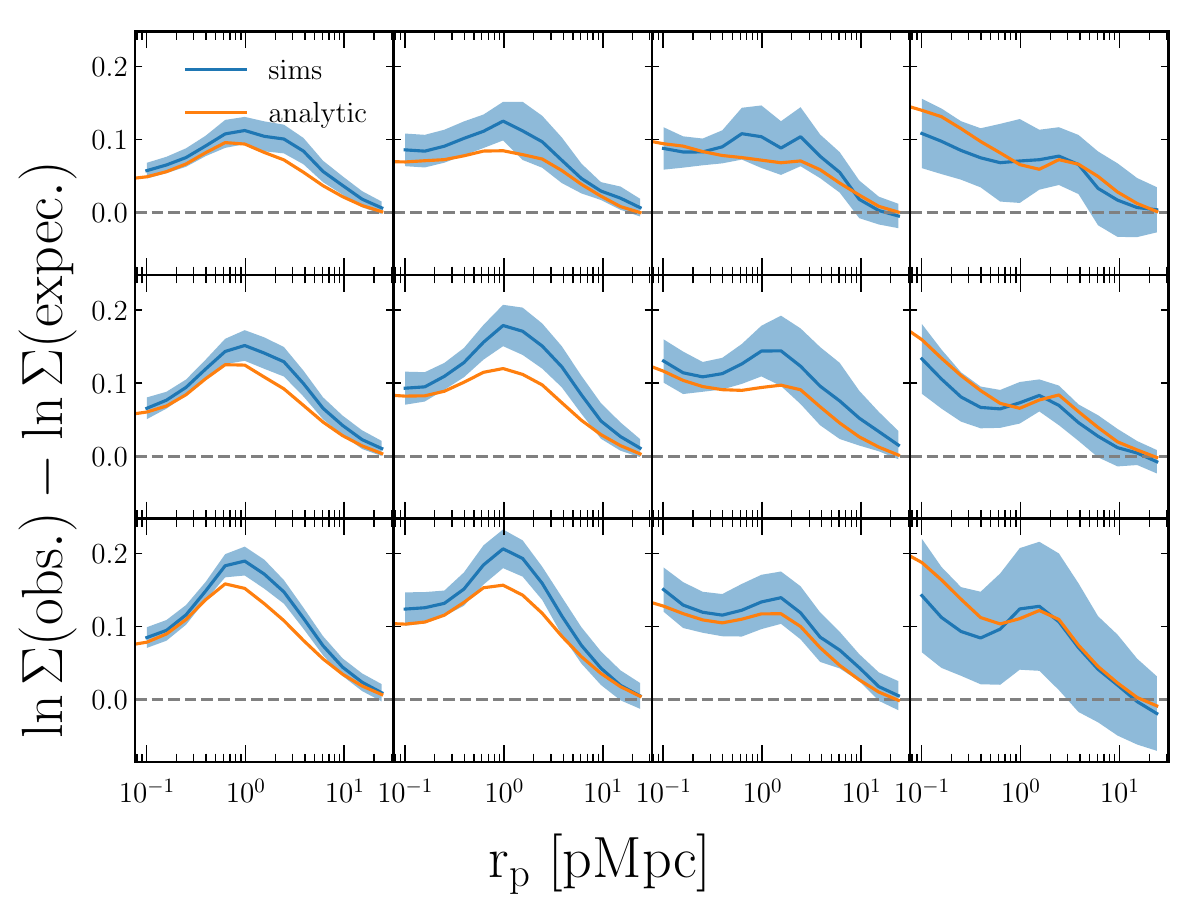}
\caption[]{Lensing selection bias estimated from equation~(\ref{eq:bias_corr}), shown in orange, based on the correlated residuals between richness and lensing at a given mass.  The blue bands show the simulation results (same as Fig.~\ref{fig:bias_DS}).}
\label{fig:bias_corr_formula}
\end{figure}
%%%%%%%%%%%%%%%%%%%%

We compare the simple model presented in equation~(\ref{eq:bias_corr}) with the selection bias results from simulations.  We first calculate the standard deviations ($\sigL$, $\sigS$) and correlation coefficient ($r$) from simulations.  To capture their mass-dependence, we use the Kernel Localized Linear Regression method \citep{KLLR}.  In this algorithm, each halo is assigned a Gaussian kernel centred on its $\lnM$; for a given mass, the linear regression is performed with each halo weighted by this kernel.  We combine all 13 Buzzard realisations, split haloes above $10^{13}\hiMsun$ into 20 log-mass bins, and use redshift bins $\Delta z = 0.15$.    We choose a Gaussian kernel width of 0.2, perform the regression independently for each $\rp$, and use 100 bootstrap samples to estimate the error bars.

The left-hand panel of Fig.~\ref{fig:corr} shows $\sigS$ as a function of projected radius, for three halo masses.  The scale-dependence is non-monotonic and exhibits a peak at $\approx$ 2 pMpc.  The right-hand panel shows the correlation between lensing residual and richness residual at a given mass, $r = {\rm corr}[\lnL - \avg{\lnL | M}, \lnS - \avg{\lnS | M}]$. For $5\times10^{13}$ and $10^{14}\ \hiMsun$, the correlation peaks at $\approx$ 0.4 pMpc, while for $5\times10^{14}\ \hiMsun$, the correlation is the largest at small radii.  From equation~(\ref{eq:bias_corr}), we can see that the selection bias is proportional to the product of $\sigS$ and $r$, and the scale-dependence we see in Fig.~\ref{fig:bias_DS} can be explained by the scale-dependence shown here.

Fig.~\ref{fig:bias_corr_formula} shows that the prediction from equation~(\ref{eq:bias_corr}) agrees well with the selection bias shown in Fig.~\ref{fig:bias_DS}.  In this calculation, we first apply equation~(\ref{eq:bias_corr}) to each halo and then average over all haloes in a given richness--redshift bin. The small discrepancy is due to the deviations from the Gaussian assumption associated with equation~(\ref{eq:bias_corr}).

To improve our understanding of selection bias, it is essential to calibrate each component in equation~(\ref{eq:bias_corr}) using simulations and observations. The scatter of the lensing signal at a given mass can be estimated from simulations. However, in the absence of an accurate galaxy--halo connection model, the correlated residuals between lensing and richness need to be modelled empirically.  One option is to use simulations to motivate a functional form for the scale-dependent correlation between the residuals and use observed lensing profiles to constrain this correlation.  In addition, one may use multi-wavelength observations to calibrate such a correlation; for example, by studying lensing and richness for a sample selected based on the Sunyaev--Zeldovich (SZ) effect.

%%%%%%%%%%%%%%%%%%%%
%%%%%%%%%%%%%%%%%%%%
\section{Projection effects and selection bias}
\label{sec:cylinder}

%%%%%%%%%%%%%%%%%%%%
\begin{figure*}
\includegraphics[width=2\columnwidth]{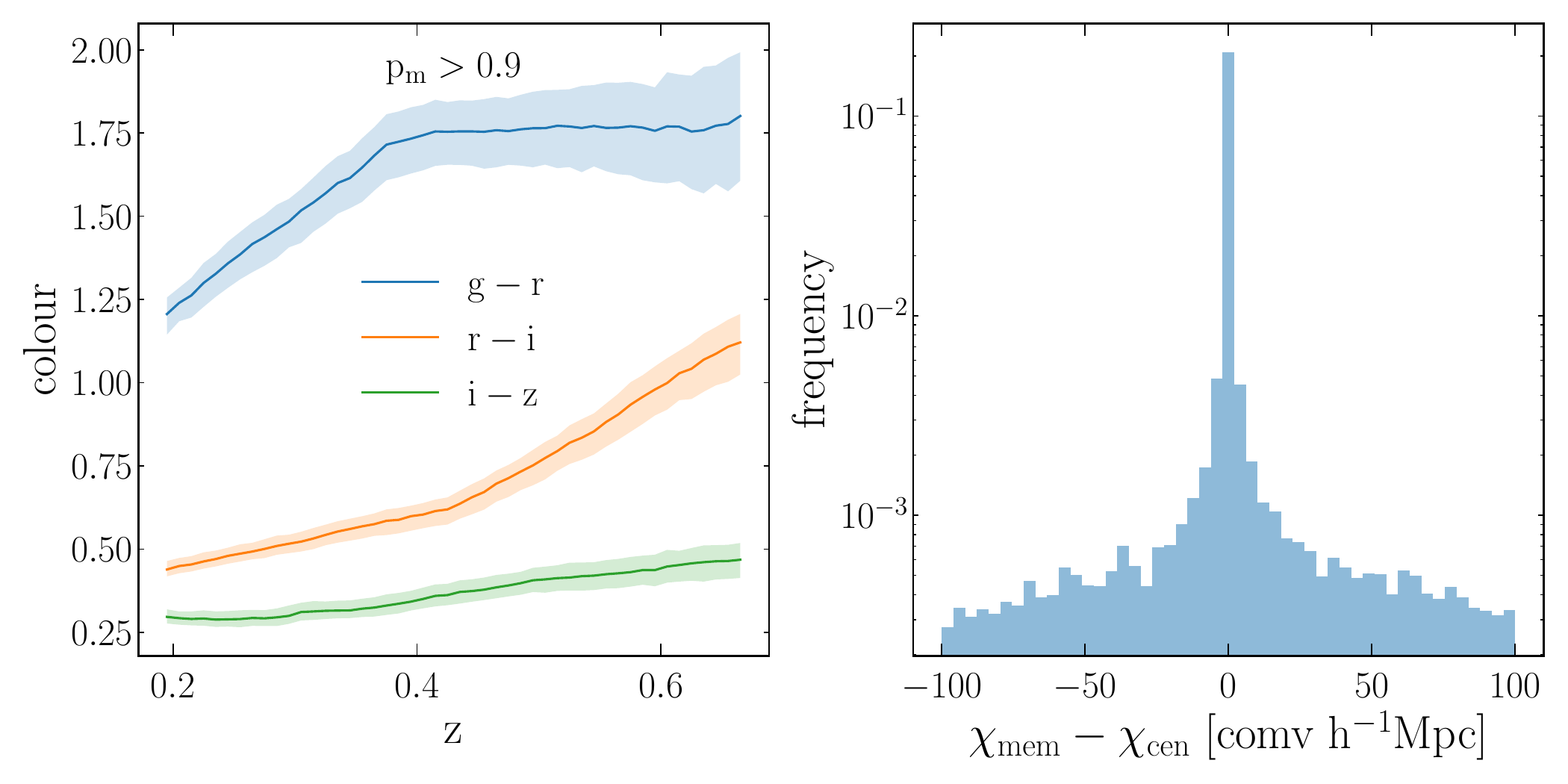}
\caption[]{Left: the colour of the \redmapper member galaxies in Buzzard as a function of cluster redshift.  The lines and bands show the medians and the 68\% intervals.  Right: the line-of-sight comoving distances between member galaxies and their host haloes.  Here we consider galaxies with a membership probability greater than 0.9.}
\label{fig:member_properties}
\end{figure*}
%%%%%%%%%%%%%%%%%%%%
\begin{figure*}
\includegraphics[width=2\columnwidth]{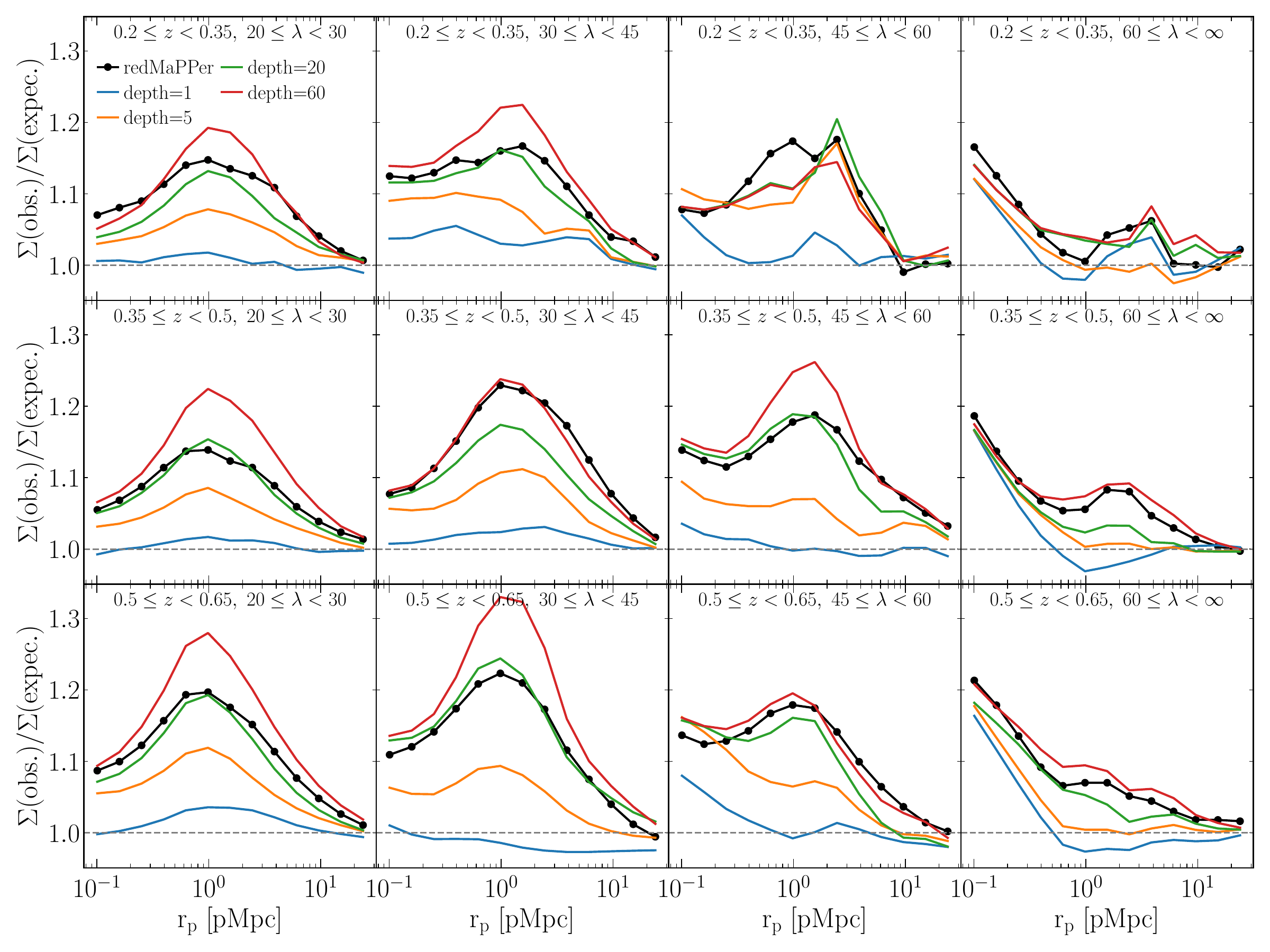}
\caption[]{Lensing selection bias from our simplified richness proxies calculated with cylinders of various projection depths.  A projection depth of $\pm 20$ to 60~$\hiMpc$ reproduces the selection bias seen in the full \redmapper calculation. For high-richness clusters, part of the selection bias is contributed by galaxies within $\pm 1 \ \hiMpc$.}
\label{fig:bias_Sigma_alt}
\end{figure*}
%%%%%%%%%%%%%%%%%%%%

We investigate the origin of selection bias by examining the impact of line-of-sight projection on the \redmapper richness.  Since \redmapper is a rather complex algorithm, we adopt a simplified approach:  we use a basic colour--magnitude cut on galaxies and count the number of galaxies in a cylinder along the line of sight.  Below we show that our simplified calculation can largely account for the lensing selection bias seen in Buzzard. In this calculation, we focus on the single DES Y3 realisation because its red-sequence width is closer to that observed by DES.

%%%%%%%%%%%%%%%%%%%%
\subsection{Properties of the \redmapper member galaxies}

We first quantify the \redmapper member galaxy properties in our simulation.  We focus on galaxies with membership probability greater than 0.9, as they form a tight red sequence and contribute to most of the richness of a cluster.

{\bf Magnitude.} We have verified that the magnitude selection of our simulated \redmapper member galaxies is consistent with the 0.2$L_*$ threshold in i-band calibrated from SDSS, as presented in equation (9) in \cite{Rykoff14}.  We use this magnitude threshold for our fiducial calculation.

{\bf Colour.} The left-hand panel of Fig.~\ref{fig:member_properties} shows the medians and 68\% intervals of the member colours (g--r, r--i, i--z), as a function of cluster redshift.  At a given   redshift, we find that the colour of red-sequence galaxies is approximately independent of  the magnitude, and we use the  median colour $\avg{c}$ and 68\% interval $\sigma_{c}$ as a simplified red-sequence template for selecting member galaxies.

For the calculations in this section, 
we select galaxies with 
\beq
\chisqc = \sum_c (c - \avg{c})^2/\sigma_c^2 < 9  \ , 
\label{eq:chisq}
\eeq
assuming no correlation between colours.  Fig.~\ref{fig:mag_color_cut} shows that changing this $\chisqc$ threshold has little effect on the resulting lensing selection bias.

{\bf Line-of-sight distances.}
We calculate the line-of-sight comoving distances between member galaxies and their host haloes.  The right-hand panel of Fig.~\ref{fig:member_properties} shows the histogram for these distances for clusters between $0.2<z<0.65$.  The prominent peak near 0 corresponds to the galaxies physically associated with the cluster, while the broad tails on both sides correspond to the projected members.   The histogram shows a transition from cluster galaxies to background galaxies at approximately 30 -- 50 $\hiMpc$.  A 50 $\hiMpc$ line-of-sight distance includes approximately 90\% of the members.  We will test the impact of projection depth in the calculations below.

%%%%%%%%%%%%%%%%%%%%
\subsection{Selecting member galaxies in a cylinder: impact of projection depth}

We calculate the number of galaxies within a cylinder around a halo centre as our mock richness.  We use galaxies from the Gold catalogue derived from Buzzard, which represents the parent galaxy sample for the DES cosmology analyses  \citep{Drlica-Wagner18}.  We use galaxies' 3D positions, magnitudes, and colours. We adopt the following fiducial choice: an aperture $\Rlambda$ (from the \redmapper output), a magnitude threshold of 0.2$L_*$ in i-band, and a colour cut of $\chisqc < 9$.  In addition, we mimic the percolation process in \redmapper to avoid double counting member galaxies: if a galaxy falls in the cylinder of multiple haloes, it is only counted as the member of the most massive halo.  We have found that including this process only slightly changes the results in the lowest richness bin.

In this section, we vary the depth of the cylinder from $\pm 1$ to 60 $\hiMpc$.  In Appendix~\ref{app:alternatives} we compare different magnitude thresholds, colour thresholds, and apertures.  For each of our cylinder-richness definitions, we use the same redshift bins as the \redmapper calculation and split the cylinder-richness into four bins, each of which has the same number of clusters as the \redmapper $\lambda$ bins.

Fig.~\ref{fig:bias_Sigma_alt} shows the selection bias in $\Sigma$ associated with different projection depths.  For all panels, a projection depth of 20 -- 60 $\hiMpc$ gives a selection bias comparable to that in the \redmapper sample (black curves).  For a projection depth of 1 $\hiMpc$, we see no selection bias for low-richness clusters but a significant selection bias for high-richness clusters for $\rp <$ 1 pMpc. This selection bias exists even when we eliminate the colour and magnitude selection. This is consistent with the $\rho(r)$ bias shown in Fig.~\ref{fig:bias_rho} and with the concentration bias shown in Fig.~\ref{fig:cvir_pdf}.  In Buzzard, for haloes above $\approx 5\times 10^{14}\ \hiMsun$, high-concentration haloes tend to have higher richness. This is opposite to the theoretical expectation that high-concentration haloes tend to form earlier and have fewer surviving satellite galaxies \citep[e.g.][]{Wu13b, Mao15}. This correlation between concentration and richness could be spurious and could lead to overestimated selection bias for high-richness clusters in Buzzard.

Given that the selection bias of \redmapper is well approximated by selecting member galaxies in a cylinder, we can use this approach to systematically study cluster selection bias in a wide range of simulations.  Since \redmapper self-calibrates the red sequence,  it can only be applied to mock catalogues with realistic galaxy colour; therefore, \redmapper has only been applied to a limited number of mock catalogues.  Our cylinder selection, on the other hand, can be readily applied to any mock galaxy catalogue for red galaxies and is computationally inexpensive. In our upcoming work, we plan to use this tool to study a wide range of mock galaxy catalogues, including those generated from hydrodynamic simulations and HOD models.

%%%%%%%%%%%%%%%%%%%%
%%%%%%%%%%%%%%%%%%%%
%%%%%%%%%%%%%%%%%%%%
\section{Discussion}
\label{sec:discussion}

In this section, we compare our results with previous work and discuss mitigation strategies for selection bias.

\subsection{Comparison with previous studies}

\citetalias{Sunayama20} quantify the impact of projection effects on cluster lensing and clustering using HOD-based galaxy catalogues.
They simulate the \redmapper richness using cylinders of depth $\pm 60 \ \hiMpc$.  We have shown that the richness calculated from $\pm 20$ to 60~$\hiMpc$ cylinders gives results broadly consistent with \redmapper; therefore, their calculation procedures and ours are comparable.  Overall, they find negligible selection bias at small scales, and their selection bias amplitude is lower than ours at large scales by approximately a factor of 2 (their Fig.~6).

We attribute the difference between their results and ours to how galaxies are populated in N-body simulations.  \citetalias{Sunayama20} assign galaxies to resolved haloes, while Buzzard assigns galaxies to both halo centres and dark matter particles.  Therefore, Buzzard has a larger field galaxy population.  In addition, \citetalias{Sunayama20} use an observationally constrained HOD, while Buzzard has lower galaxy number densities in cluster-size haloes compared with observations, resulting in lower cluster number counts above a given richness (see Fig.~4 and Fig.~8 in \citealt{Wechsler21} and Fig.~12 and Fig.~13 in \citealt{DeRose19}). The combination of a lower cluster galaxy content and a larger background population compared with observations leads to stronger projection effects in Buzzard.

Several analyses also suggest that the lensing selection bias in Buzzard is higher than that in observations.  In \citetalias{DESY1CL}, the lensing selection bias needed to reconcile cluster counts with DES 3$\times$2pt analysis is weaker than what we have calculated from Buzzard (their Fig.~12).  Combining cluster abundances, cluster lensing and clustering, and galaxy lensing and clustering, \cite{To21b} find that the best-fitting selection bias in DES Y1 clusters is $\approx 15\%$ for scales $\gtrsim 8 \ \hiMpc$ (comoving), which is smaller than that in Buzzard.

Because of the uncertainty associated with galaxy modelling, care must be taken when applying lensing selection bias derived from simulations to cosmological analyses. A conservative approach is to adopt a functional form for the selection bias motivated by simulations and let the observational data  self-calibrate the model parameters. This is similar to the approach in \cite{To21b}, who model the large-scale selection bias with a power law in mass and fit for the model parameters \citep[also see][]{Park21}.  However, in an analysis combining  cluster counts and weak lensing such as \citetalias{DESY1CL}, fully uninformative priors on the selection bias parameters would completely wash out the cosmological constraining power of the catalogue.  It is thus necessary to develop informative priors using a wide range of galaxy models, which could be achieved by exploring HOD parameters.

An alternative approach is to model the correlated residuals between observables and fit for the correlation coefficients \citep[e.g.][]{Grandis21a, Chiu21}.  This approach extends the analytic model described in Section~\ref{sec:basics} and can be part of the forward-modelling procedure.  However, this approach could lead to many  weakly constrained nuisance parameters, and informative priors on the correlation coefficients would also be necessary.

It would be valuable to develop emulators for calculating  the selection bias for a range of galaxy models and cluster selection methods.  Recent studies have used N-body simulation-based emulators to predict cluster lensing on non-linear scales \citep[e.g.][]{Nishimichi19,Salcedo20,Cromer21}, but they have so far used halo-based cluster selection. With cylinder selection as a proxy for the full \redmapper selection, it may be feasible to construct emulators that directly model the cluster selection procedure.

Our results indicate that the \redmapper cluster projection effects are dominated by galaxies within approximately $\pm 60\ \hiMpc$ along the line-of-sight, which corresponds to a redshift difference of 0.02.  This is slightly smaller than the observational results in \cite{Myles20projection}.  They fit the spectroscopic redshift distribution of galaxies associated with \redmapper clusters using a double Gaussian mixture model to account for true and spurious members, finding that the latter has a standard deviation of approximately 0.03.  Their result is consistent with \cite{Costanzi19projection}, who develop a projection effect proxy $\sigma_z$ to describe how widely cluster members are spread along the line of sight. We plan to use these observations to constrain the projection depth and to improve the modelling of projection effects in simulations.

Baryonic effects have been shown to alter the cluster lensing signal and bias the lensing-derived mass low by 5 to 10\% \citep[e.g.][]{Bahe12, Henson17, Lee18, Debackere21, Grandis21b}; however, selection bias is comparable or larger in magnitude, extends to larger scales, and is potentially more difficult to model because it depends on the uncertain  relation between galaxies and haloes.  Therefore, we expect that projection effects would be the most significant systematic uncertainty in optical cluster lensing.  On the other hand, baryonic effects tend to make clusters more spherical \citep[e.g.][]{Bryan13, Henson17}, which would reduce orientation bias.  It would be valuable to use full hydrodynamic simulations with reliable galaxy populations to self-consistently study the projection-induced correlated residuals between lensing and richness.

Forecast studies show that the cosmological precision attainable from cluster weak lensing is competitive with that attainable from cosmic shear analyses of the same weak lensing data set, if the statistical limits can be achieved \citep[e.g.][]{OguriTakada11, YooSeljak12, Weinberg13, Salcedo20, Wu21}.  For example, \cite{Wu21} find that a DES-like survey of cluster lensing could achieve 0.26\% precision on $\sigma_8$ (with other cosmological parameters held fixed) if the mass--observable scatter is constrained independently, and \cite{Salcedo20} forecast a $\sigma_8$ precision of better than 1\% if the scatter is not known independently but constrained by cluster--galaxy cross-correlations and galaxy auto-correlations. The challenge is to realise this statistical precision in the face of selection bias that affects $\Sigma(\rp)$ at the 10--20\% level.

\subsection{Mitigation strategies}

Below we discuss strategies for mitigating the cluster lensing selection bias.

{\bf Simulating different galaxy models.}
As discussed earlier, the lensing selection bias depends on the underlying galaxy population.  One way to reduce this modelling uncertainty is to quantify how projection effects depend on the HOD parameters of galaxies that contribute to the \redmapper richness.  Specific choices of HOD parameters have been studied (e.g. \citealt{Costanzi19projection}; \citetalias{Sunayama20}), but quantifying their effects would require a systematic study of a wide range of HOD models.  For example, HOD models with a larger satellite fraction or more galaxies in low-mass haloes would exhibit stronger projection effects.

{\bf Combining multi-wavelength cluster observables.}
We can potentially use multi-wavelength observations to quantify the optical selection bias. For example, one can use the SZ signal to select clusters and study the correlated residuals between richness and lensing, taking advantage of the small mass scatter of SZ-selected clusters.  Such cluster samples exist for $\lambda \gtrsim 50$ systems and have been used to calibrate the scatter of optical clusters \citep[e.g.][]{RozoRykoff14, Saro15, Farahi19}. In addition, the cross-comparison between optical and SZ clusters can be used to study projection effects \citep[e.g.][]{Grandis21a}; for example, a cluster heavily contaminated by galaxies along the line of sight would have a lower SZ signal than expected from its richness.

{\bf Combining cluster lensing and clustering.} 
The multi-wavelength approach described above is usually not applicable to low-richness clusters.  To calibrate the selection bias for low-richness clusters, one can combine the clustering and lensing of galaxy clusters. At large scales, the selection bias manifests as the clustering bias, and we can calibrate it by combining cluster lensing, cluster--galaxy cross-correlation, and galaxy auto-correlation.  This is similar to combining clustering and lensing to constrain the mass--observable relation \citep[e.g.][]{Salcedo20, Chiu20, To21b}.

{\bf Spectroscopic observations} of member galaxies can distinguish true from spurious member galaxies \citep[e.g.][]{Rozo15RM4, Sohn18, Rines18, Myles20projection, Wetzell21}.  As we have shown in Section~\ref{sec:cylinder}, galaxies with different line-of-sight distances lead to different amounts of selection bias.  Therefore, quantifying the redshift distribution of the \redmapper member galaxies associated with the line-of-sight structure would reduce the modelling uncertainties of selection bias.  The Dark Energy Spectroscopic Instrument (DESI) and Roman Space Telescope's grism spectroscopy will provide large cluster samples for such analyses.

{\bf Stellar mass} can potentially serve as a low-scatter mass proxy, especially with optimally chosen member galaxies \cite[e.g.][]{Golden-Marx18, Bradshaw20, Anbajagane20, Huang21}. A cluster sample selected by stellar mass could also provide a useful sanity check for the commonly used richness selection   \citep[e.g.][]{Pereira18, Pereira20, Palmese20}.  The stellar mass may have weaker projection effects than richness because it has a large contribution from the brightest cluster galaxies.

{\bf Defining a cluster sample by a threshold.}
One way to simplify the modelling of selection bias is to use a single richness threshold instead of multiple richness bins to define our cluster sample.  \cite{Wu21} have shown that the former requires fewer nuisance parameters for the mass--observable relation and can avoid diluting the cosmological information.  In addition, the threshold approach does not require the power-law assumption of the mass--observable relation, an assumption that could be too restrictive. Similarly, the threshold approach only requires modelling the selection bias near the richness threshold and can significantly simplify the analyses.

Mitigating selection bias requires us to consider all aspects of the cosmological parameter inference, including the impact of projection on cluster number counts and on the determination of the richness--mass scatter, which is the critical nuisance parameter for analyses focusing on cluster number counts and weak lensing. For example, in a conventional analysis such as \citetalias{DESY1CL}, which models number counts and $\DS(\rp)$ in richness and redshift bins based on the halo mass function and a parameterised richness--mass relation, one can incorporate selection bias curves like those in Fig.~\ref{fig:bias_DS} into the model prediction.  It is essential to study a wider range of galaxy HODs and cosmologies to establish the appropriate priors for such corrections.  Other approaches bring in cluster auto-correlations, cluster--galaxy cross-correlations, and galaxy auto-correlations as additional constraints \citep{Salcedo20, To21a}, and for these one must examine the impact of selection bias on these additional observables.

%%%%%%%%%%%%%%%%%%%%
%%%%%%%%%%%%%%%%%%%%
\section{Summary}
\label{sec:summary}

We investigate the bias of the stacked weak lensing signals around optically selected clusters, using the \redmapper cluster finder applied to the Buzzard simulations.  We find that the large-scale excess surface mass density $\Delta\Sigma(\rp)$ of richness-selected clusters in Buzzard is 20 -- 60\% higher than that expected from the underlying halo mass PDF. Expressed in surface mass density $\Sigma(\rp)$ rather than $\Delta\Sigma(\rp)$, the bias shows strong scale-dependence and peaks at $\rp\approx 1$ pMpc with an amplitude of 10 -- 20\%.  This scale-dependence is well explained by an analytical model that accounts for the correlated residuals between the surface mass density and richness at a given halo mass (equation~\ref{eq:bias_corr}).  The correlated residuals arise mainly from projection effects, the boosting of richness and surface mass density by galaxies and matter that lie along the line of sight but outside the halo virial radius.  At high richness and small scales, the preferential selection of higher concentration haloes also makes a significant contribution.

We have shown that the complex \redmapper cluster selection can be modelled by a cylinder member selection. We have found that galaxies within $\pm 20$ to 60~$\hiMpc$ along the line of sight but outside the halo virial radius are the main cause of the selection bias.  This simplified cylinder selection method can be efficiently applied to a wide range of simulations to study the impact of the galaxy model on selection bias.  Our ultimate goal is to mitigate the impact of this bias on cosmological constraints derived from cluster weak lensing surveys.

The selection bias is currently one of the key systematic effects that limit the statistical power of optical cluster cosmology analyses.  As discussed above, solving the selection bias would require a concerted effort of simulations, multi-wavelength observations, and combined-probe analyses. Currently, DES provides an unprecedented data set for cluster weak lensing, and in the next decade Euclid, LSST, and the Roman Space Telescope will all provide data sets that are more powerful still. Exploiting the measurements from these data sets is a theoretical challenge, with a potentially critical payoff in unveiling the physics behind cosmic acceleration.

\section*{Acknowledgements}

HW is supported by DOE Grant DE-SC0021916 and NASA grant 15-WFIRST15-0008.
MESP is funded by the Deutsche Forschungsgemeinschaft (DFG, German Research Foundation) under Germany's Excellence Strategy -- EXC 2121 `Quantum Universe' -- 390833306.

This research uses resources of the National Energy Research Scientific Computing Center (NERSC), a U.S.~Department of Energy Office of Science User Facility located at Lawrence Berkeley National Laboratory.
HW acknowledges the high-performance computing support of the Borah computer cluster (DOI: 10.18122/oit/3/boisestate) provided by Boise State University's Research Computing Department.

This paper has gone through an internal review by the DES collaboration.

Funding for the DES Projects has been provided by the U.S. Department of Energy, the U.S. National Science Foundation, the Ministry of Science and Education of Spain, 
the Science and Technology Facilities Council of the United Kingdom, the Higher Education Funding Council for England, the National Center for Supercomputing 
Applications at the University of Illinois at Urbana-Champaign, the Kavli Institute of Cosmological Physics at the University of Chicago, 
the Center for Cosmology and Astro-Particle Physics at the Ohio State University,
the Mitchell Institute for Fundamental Physics and Astronomy at Texas A\&M University, Financiadora de Estudos e Projetos, 
Funda{\c c}{\~a}o Carlos Chagas Filho de Amparo {\`a} Pesquisa do Estado do Rio de Janeiro, Conselho Nacional de Desenvolvimento Cient{\'i}fico e Tecnol{\'o}gico and 
the Minist{\'e}rio da Ci{\^e}ncia, Tecnologia e Inova{\c c}{\~a}o, the Deutsche Forschungsgemeinschaft and the Collaborating Institutions in the Dark Energy Survey. 

The Collaborating Institutions are Argonne National Laboratory, the University of California at Santa Cruz, the University of Cambridge, Centro de Investigaciones Energ{\'e}ticas, 
Medioambientales y Tecnol{\'o}gicas-Madrid, the University of Chicago, University College London, the DES-Brazil Consortium, the University of Edinburgh, 
the Eidgen{\"o}ssische Technische Hochschule (ETH) Z{\"u}rich, 
Fermi National Accelerator Laboratory, the University of Illinois at Urbana-Champaign, the Institut de Ci{\`e}ncies de l'Espai (IEEC/CSIC), 
the Institut de F{\'i}sica d'Altes Energies, Lawrence Berkeley National Laboratory, the Ludwig-Maximilians Universit{\"a}t M{\"u}nchen and the associated Excellence Cluster Universe, 
the University of Michigan, NSF's NOIRLab, the University of Nottingham, The Ohio State University, the University of Pennsylvania, the University of Portsmouth, 
SLAC National Accelerator Laboratory, Stanford University, the University of Sussex, Texas A\&M University, and the OzDES Membership Consortium.

Based in part on observations at Cerro Tololo Inter-American Observatory at NSF's NOIRLab (NOIRLab Prop. ID 2012B-0001; PI: J. Frieman), which is managed by the Association of Universities for Research in Astronomy (AURA) under a cooperative agreement with the National Science Foundation.

The DES data management system is supported by the National Science Foundation under Grant Numbers AST-1138766 and AST-1536171.
The DES participants from Spanish institutions are partially supported by MICINN under grants ESP2017-89838, PGC2018-094773, PGC2018-102021, SEV-2016-0588, SEV-2016-0597, and MDM-2015-0509, some of which include ERDF funds from the European Union. IFAE is partially funded by the CERCA program of the Generalitat de Catalunya.
Research leading to these results has received funding from the European Research
Council under the European Union's Seventh Framework Program (FP7/2007-2013) including ERC grant agreements 240672, 291329, and 306478.
We  acknowledge support from the Brazilian Instituto Nacional de Ci\^encia
e Tecnologia (INCT) do e-Universo (CNPq grant 465376/2014-2).

This manuscript has been authored by Fermi Research Alliance, LLC under Contract No. DE-AC02-07CH11359 with the U.S. Department of Energy, Office of Science, Office of High Energy Physics.

\section*{Data Availability}

The data used in this work is available upon request.

\bibliographystyle{mnras}
\bibliography{master_refs}

\appendix
%%%%%%%%%%%%%%%%%%%%
%%%%%%%%%%%%%%%%%%%%
\section{Comparing two versions of the Buzzard simulations}\label{app:buzzard_versions}

%%%%%%%%%%%%%%%%%%%%
\begin{figure}
\centering
\includegraphics[width=1\columnwidth]{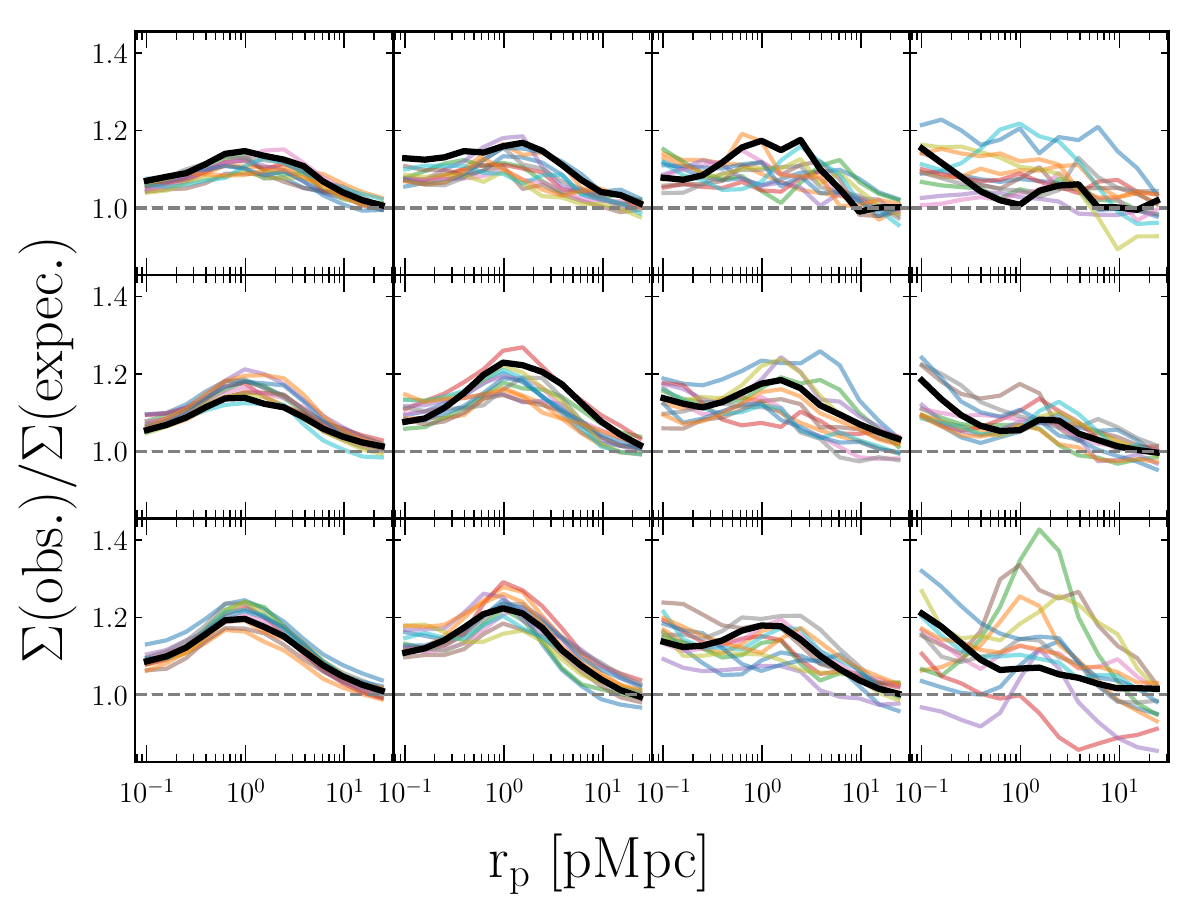}
\caption[]{Comparison between two versions of Buzzard and different realisations.  The colour curves correspond to the 12 DES Y1 realisations, and the black curve corresponds to the single DES Y3 realisation. The statistical fluctuation is larger than the difference between the two versions.  Therefore, in the main text, we show the area-weighted means and standard deviations of all 13 realisations.}
\label{fig:bias_DS_buzzard_versions}
\end{figure}
%%%%%%%%%%%%%%%%%%%

As described in Section~\ref{sec:sims}, we use two versions of Buzzard:  12 realisations of the DES Y1 data (1120 deg$^2$) based on Buzzard version 1.9.2+1 \citep[presented in][]{DeRose19}, and one realisation of the DES Y3 data (4143 deg$^2$) based on Buzzard version 1.9.8 \citep[presented in][]{DeRose22}. The main differences between the two versions are as follows:

\begin{itemize}

    \item For the subhalo abundance matching, the former uses the DES Y1 luminosity functions, while the latter uses the DES Y3 luminosity functions (both are modified from the SDSS luminosity functions).
    
    \item For the environmental proxy for the SED assignment, the former uses a galaxy's projected distance to the fifth nearest neighbour, while the latter uses a galaxy's 3D distance to the nearest halo above a given mass.
    
    \item The former has a narrower red sequence compared with DES data, while the latter explicitly matches the mean and scatter of the red sequence observed in DES Y3.
    
\end{itemize}

Fig.~\ref{fig:bias_DS_buzzard_versions} compares the lensing bias derived from the two versions and their individual realisations: the colour curves show the 12 DES Y1 realisations, and the black curve shows the one DES Y3 realisation.  Overall, the statistical fluctuations associated with different realisations are larger than the difference between the two versions.   While \cite{To21a} found that the galaxy clustering of realisation `3b' is problematic and removed it from their analysis, we do not find such a discrepancy in cluster lensing.  Therefore, unless otherwise noted, throughout this paper we combine all 13 realisations and calculate area-weighted means and standard deviations.

%%%%%%%%%%%%%%%%%%%%
%%%%%%%%%%%%%%%%%%%%
\section{Comparing diagnosis methods for lensing selection bias}\label{app:diagnosis}

%%%%%%%%%%%%%%%%%%%%
\begin{figure}
\centering
\includegraphics[width=\columnwidth]{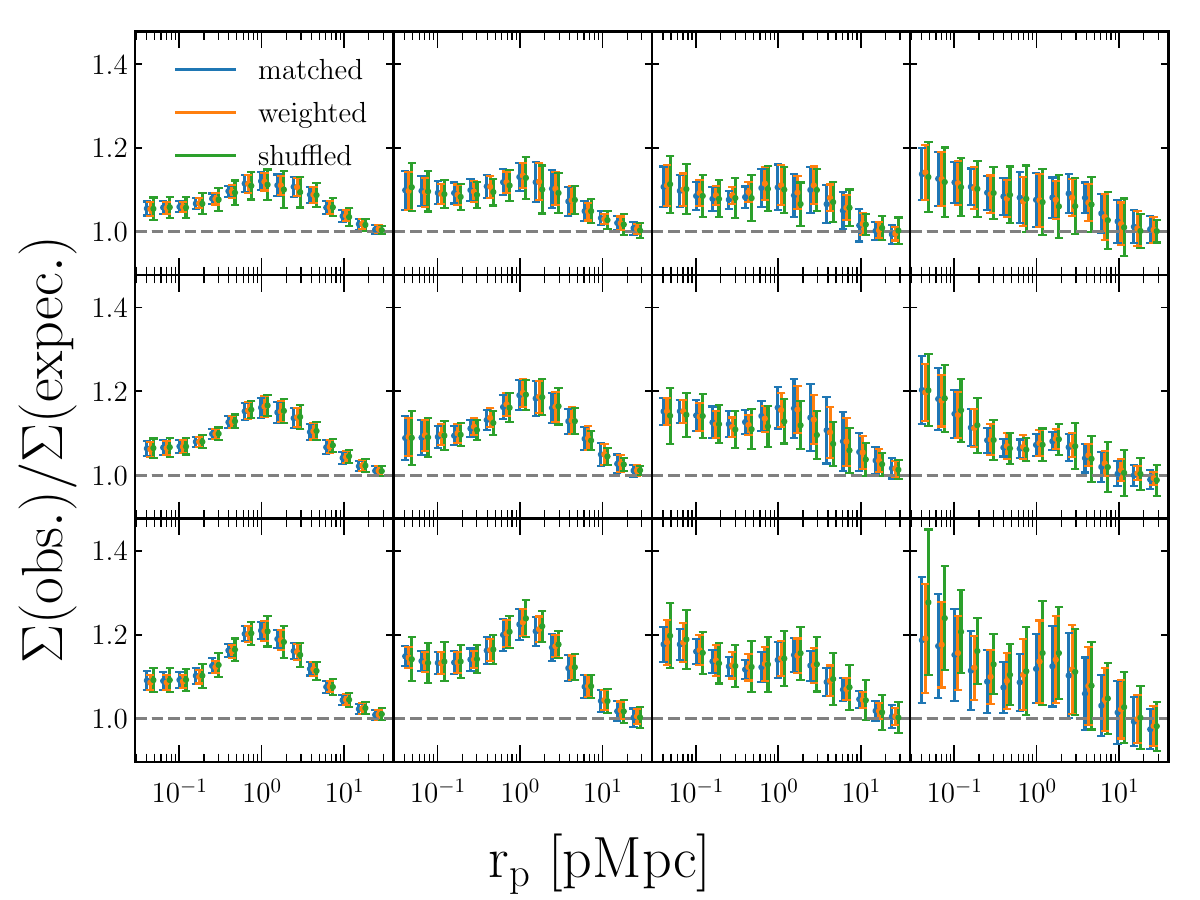}
\caption[]{Comparison between the three diagnosis methods for lensing selection bias detailed in Appendix~\ref{app:diagnosis}.  The three methods agree well with each other.}
\label{fig:bias_DS_diagnosis}
\end{figure}
%%%%%%%%%%%%%%%%%%%% 

Here we compare three diagnosis methods for cluster lensing selection bias, which is quantified by the ratio between (1) the lensing signal from a sample selected in a richness and redshift bin, similar to the DES cluster analyses (we call this the `richness-selected' sample), and (2) the lensing signal we expect from the underlying halo mass PDF of the richness-selected sample. 
The former is straightforward to compute. Below we present three methods to compute the latter. For all cases, we split each redshift bin $\Delta z = 0.15$ into 3 narrower bins $\Delta z = 0.05$ to account for the redshift dependence. 

\begin{enumerate}

\item {\bf Shuffling richness at a given halo mass.}
We start with the full halo catalogue and put haloes in narrow mass bins.  For haloes in a given mass bin, we shuffle their richness values.  This procedure washes out any correlated residuals between lensing and richness at a given mass.  Each halo is assigned a new richness $\lambda^{\rm shuff}$.  We calculate the ratio between the stacked lensing signal of clusters selected with $\lambda$ and that of clusters selected with $\lambda^{\rm shuff}$.  This ratio corresponds to the biased lensing signal due to the correlated residuals between richness and lensing.

\item {\bf Matching the underlying mass PDF.}
From the full halo catalogue, we select {\em random} haloes to match exactly the same mass PDF as the richness-selected sample. We construct this random sample to be 5 times the number of the richness-selected sample, although for massive haloes we need to draw with replacement.  We call this the `mass-matched' sample and calculate its mean weak lensing signal.

\item {\bf Weighting by the underlying mass PDF.} 
This method is analogous to the matching method, but instead of constructing a random halo sample to match the mass PDF of the richness-selected sample, we use all haloes in the catalogue weighted by this PDF.  We use the full halo catalogue and put haloes in narrow mass bins and calculate the mean lensing signal from this bin.  Using the PDF of the richness-selected sample, we can calculate the weight associated with each narrow mass bin.  We then perform a weighted average of the lensing signal from all mass bins.  

\end{enumerate}
%%%%%%%%%%%%%%%%%%%%

Fig.~\ref{fig:bias_DS_diagnosis} shows that all three methods give consistent results.  The shuffling method is the easiest to understand but is also the noisiest because it uses the smallest number of haloes.  The weighting method is the least noisy because it averages over the lensing of all haloes in the catalogue.  In the main text, we present the results calculated from the weighting method.  As an additional sanity check, we have constructed catalogues with no selection bias by assigning to each halo a random richness with a log-normal scatter, and we have recovered unbiased results.

%%%%%%%%%%%%%%%%%%%%
%%%%%%%%%%%%%%%%%%%%
\section{Impact of magnitude, colour, and aperture on selection bias}\label{app:alternatives}
%%%%%%%%%%%%%%%%%%%%
\begin{figure}
\includegraphics[width=1\columnwidth]{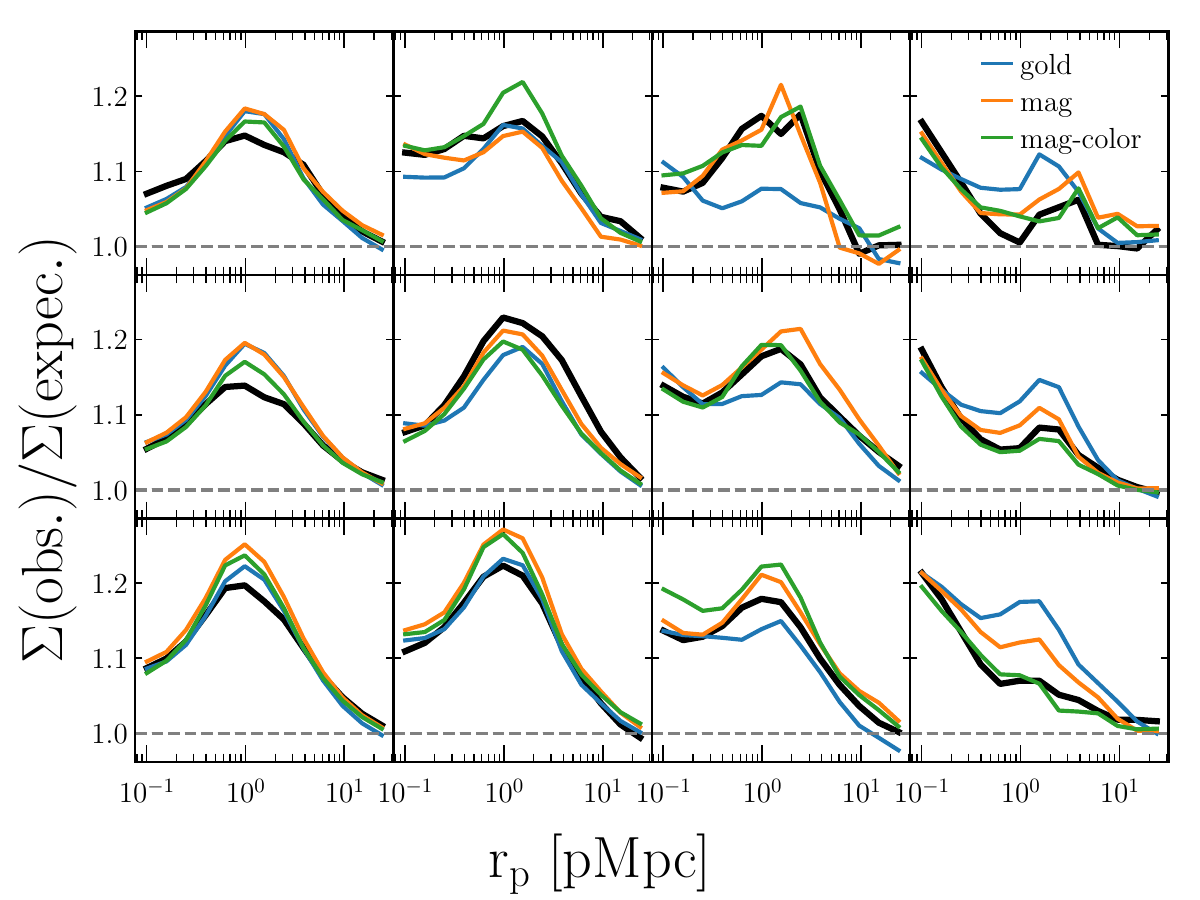}
\caption[]{Impact of member galaxy magnitude and colour cuts on lensing selection bias.  We select members using a cylinder of radius $\Rlambda$ and length $\pm 30\ \hiMpc$ and compare the richness calculated with (1) using all galaxies in the Gold catalogue, (2) setting a magnitude cut at 0.2$L_*$, and (3) setting an additional colour cut at $\chisqc < 9$.   While the magnitude and colour cuts significantly reduce the selection bias for high-richness clusters, they have weak impact on low-richness clusters.}
\label{fig:mag_color_cut}
\end{figure}
%%%%%%%%%%%%%%%%%%%%
\begin{figure}
\includegraphics[width=\columnwidth]{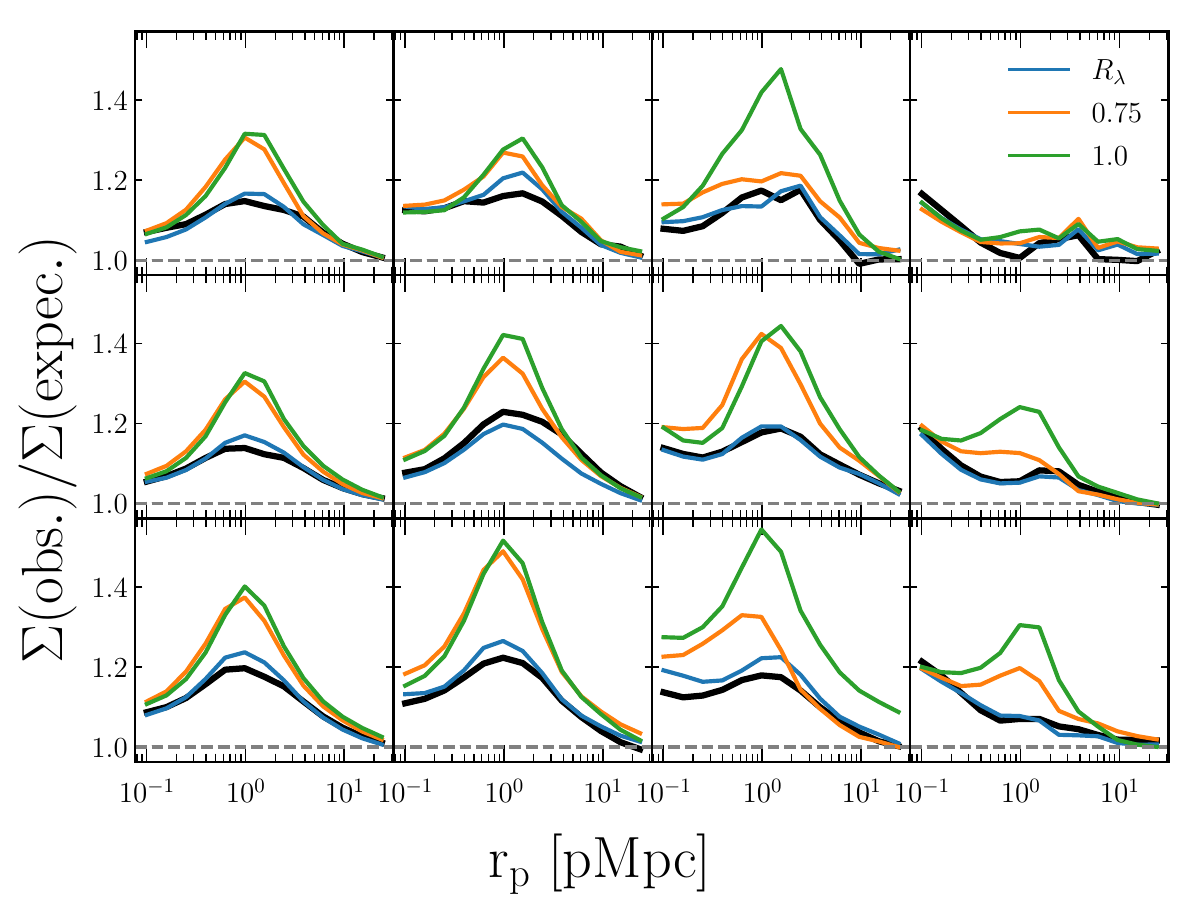}
\caption[]{Impact of member galaxy selection aperture on lensing selection bias. We compare $\Rlambda$ with fixed, richness-independent apertures: 0.75 and 1 physical $\hiMpc$.  A fixed aperture increases the selection bias.}
\label{fig:aperture}
\end{figure}
%%%%%%%%%%%%%%%%%%%%

In the main text, we present the impact of projection depth on lensing selection bias.  In this appendix, we compare different magnitude cuts, colour cuts, and apertures for selecting member galaxies.

Fig.~\ref{fig:mag_color_cut} shows the selection bias using galaxies selected with different magnitude and colour criteria.   For each cluster, we use a cylinder of radius $\Rlambda$ and depth $\pm 30\ \hiMpc$ to define its richness.  We compare the \redmapper results with (1) galaxies in the Gold catalogue, which correspond to an i-band magnitude limit of $\approx$ 26 and no colour cut, (2) a magnitude cut at 0.2$L_*$ and no colour cut, and (3) our fiducial magnitude and colour cut (0.2$L_*$ and $\chisqc < 9$, see equation~\ref{eq:chisq}).   Releasing the magnitude and colour cut increases the selection bias for high-richness clusters but has small effects on low-richness clusters.

Fig.~\ref{fig:aperture} shows the selection bias associated with richness defined by different apertures, using our fiducial magnitude--colour selection and cylinder depth $\pm 30 \ \hiMpc$ (comoving).  We compare $\Rlambda = 1 (\lambda/100)^{0.2}~\hiMpc$ (physical) with a fixed, richness-independent aperture, 0.75 and 1 physical $\hiMpc$.   We find that using a fixed aperture leads to a higher selection bias compared with $\Rlambda$.   This increased selection bias is related to the increased scatter of mass at a given richness when using a fixed aperture.  Although $\Rlambda$ and fixed apertures give very similar scatter in $\lambda$ at a given mass, the latter gives a shallower slope of $\lambda$--mass relation, which leads to a larger scatter in mass at a given $\lambda$.  \cite{Rykoff12} chose the relation between $\lambda$ and $\Rlambda$ to minimise the scatter of $L_X$ at a fixed $\lambda$, and it is encouraging that this radius also gives a weaker selection bias.

%%%%%%%%%%%%%%%%%%%%
%%%%%%%%%%%%%%%%%%%%
\section{Selection bias associated with halo orientation}
\label{app:triaxiality}

\subsection{Measuring the triaxial shape of haloes in simulations}

We use a triaxial ellipsoid to describe the 3D shape of a halo, and its orientation is described by the angle between its major axis and the line of sight, $i$.  For a halo sample with random orientations, $\cosi$ follows a uniform distribution, because the surface area element on a sphere is given by $d(\cosi) d\phi$, where $\phi$ is the azimuthal angle and runs from 0 to $2\pi$.

For haloes in Buzzard, we measure the triaxial shapes and axis orientations following the method described in \cite{Osato18} with slight modifications.  We use all dark matter particles inside $\Rvir$ to iteratively measure the reduced inertia tensor and its eigensystems.  We adopt the convention $a \leq b \leq c$ for an ellipsoid, following \cite{Osato18, JingSuto02}.  Unlike \cite{Bett12}, we do not trim particles in each iteration.  We have tested that using slightly different radii,  trimming particles in each iteration, or using a non-reduced tensor changes the $\cosi$ by less than 0.1.

The iterative calculation of the halo shape starts with 
\beqa
q &= 1 \ , \quad s = 1 \ , \\
R_{\rm p,1} &= x \ , \quad R_{\rm p,2} = y \ , \quad R_{\rm p,3} = z \ , \\
R_{\rm p}^2 &= \left(\frac{R_{\rm p,1}}{q}\right)^2 + \left(\frac{R_{\rm p,2}}{s}\right)^2 + R_{\rm p,3}^2 \ , \\
\mathcal{M}_{ij} &= \frac{1}{N_{\rm p}}\sum_{p=1}^{N_{\rm p}} \frac{R_{{\rm p},i}R_{{\rm p},j} }{R_{\rm p}^2} \quad\quad \mbox{for $i,j \in (1,2,3)$}  \ ,
\eeqa
where $(x, y, z)$ are the positions of individual particles, $N_{\rm p}$ is the number of particles, and the subscript $p$ runs through all particles.
We calculate the eigenvalues ($\lambda_1$, $\lambda_2$, $\lambda_3$), sorted from small to large, and the corresponding eigenvectors ($\boldsymbol{v}_1$, $\boldsymbol{v}_2$, $\boldsymbol{v}_3$),
\beqa
a &= \sqrt{\lambda_1},\ \boldsymbol{v}_{\rm minor} = \boldsymbol{v}_1 \\
b &= \sqrt{\lambda_2},\ \boldsymbol{v}_{\rm int} = \boldsymbol{v}_2 \\
c &= \sqrt{\lambda_3},\ \boldsymbol{v}_{\rm major} = \boldsymbol{v}_3   \ .
\eeqa
We then update the values:
\beqa
q &= a/c \\
s &= b/c \\
R_{{\rm p},1} &= \boldsymbol{R}_{\rm p} \cdot \boldsymbol{v}_1  \\
R_{{\rm p},2} &= \boldsymbol{R}_{\rm p} \cdot \boldsymbol{v}_2  \\
R_{{\rm p},3} &= \boldsymbol{R}_{\rm p} \cdot \boldsymbol{v}_3  \\
R_p^2 &= \left(\frac{R_{{\rm p},1}}{q}\right)^2 + \left(\frac{R_{{\rm p},2}}{s}\right)^2 + R_{{\rm p},3}^2   \ .
\eeqa
These numbers are used for the next iteration.

In each iteration, we transform the coordinate system using the matrix
\[ 
\mathbf{R}_i = \left [
  \begin{tabular}{c}
  $\boldsymbol{v}_1^T$\\
  $\boldsymbol{v}_2^T$\\
  $\boldsymbol{v}_3^T$ 
  \end{tabular}
\right ]  \ .
\]
We multiply the rotation matrix in each step:
\beq
\mathbf{R}_{\rm final} =  \mathbf{R}_n \cdots \mathbf{R}_1 \mathbf{R}_0  \ ,
\eeq
and the third row of $\mathbf{R}_{\rm final}$ is the major axis, denoted as $\boldsymbol{v}_3^{\rm final}$.
The orientation with respect to the line of sight is given by
\beq
\cos(i) = |\boldsymbol{v}_3^{\rm final} \cdot \boldsymbol{v}^{\rm LOS}|  \ .
\eeq
This $\cos(i)$ is very similar to the result using $\boldsymbol{v}_3$ in the initial step.
The iteration ends when the fractional changes in both $q$ and $s$ are less than $10^{-7}$.
In Buzzard, the observer is placed at the origin of the $z = 0$ N-body simulation box, and thus the line-of-sight direction is the same as the position vector.

\subsection{Orientation PDF for the Buzzard \redmapper clusters}
%%%%%%%%%%%%%%%%%%%
\begin{figure}
\includegraphics[width=1\columnwidth]{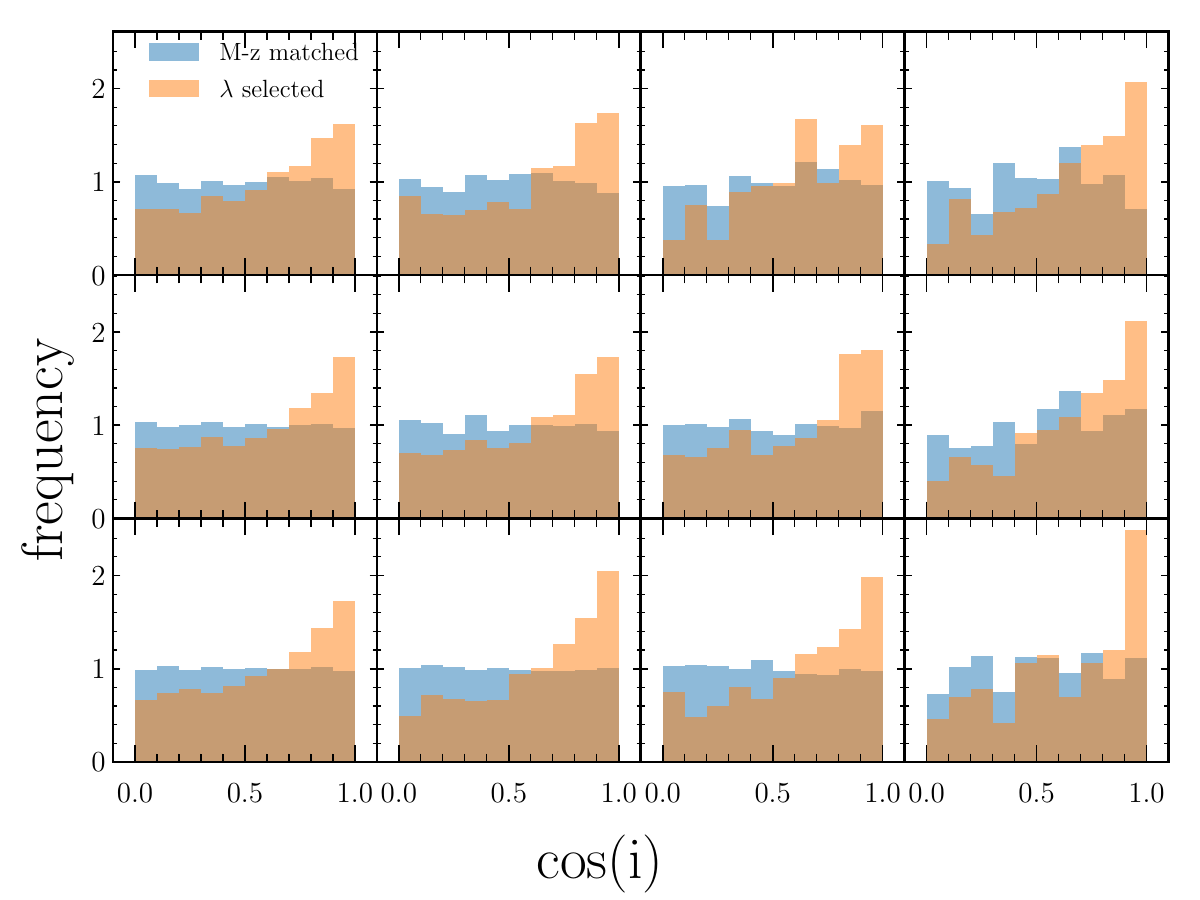}
\caption[]{Probability distribution of halo orientation $\cosi$ for the richness-selected sample (orange) and the sample constructed to match the underlying mass and redshift distribution (blue).  The panels match the redshift and richness bins in Fig.~\ref{fig:bias_DS}. The richness-selection preferentially selects haloes with high $\cosi$.}
\label{fig:cosi_pdf}
\end{figure}
%%%%%%%%%%%%%%%%%%%
\begin{figure}
\includegraphics[width=1\columnwidth]{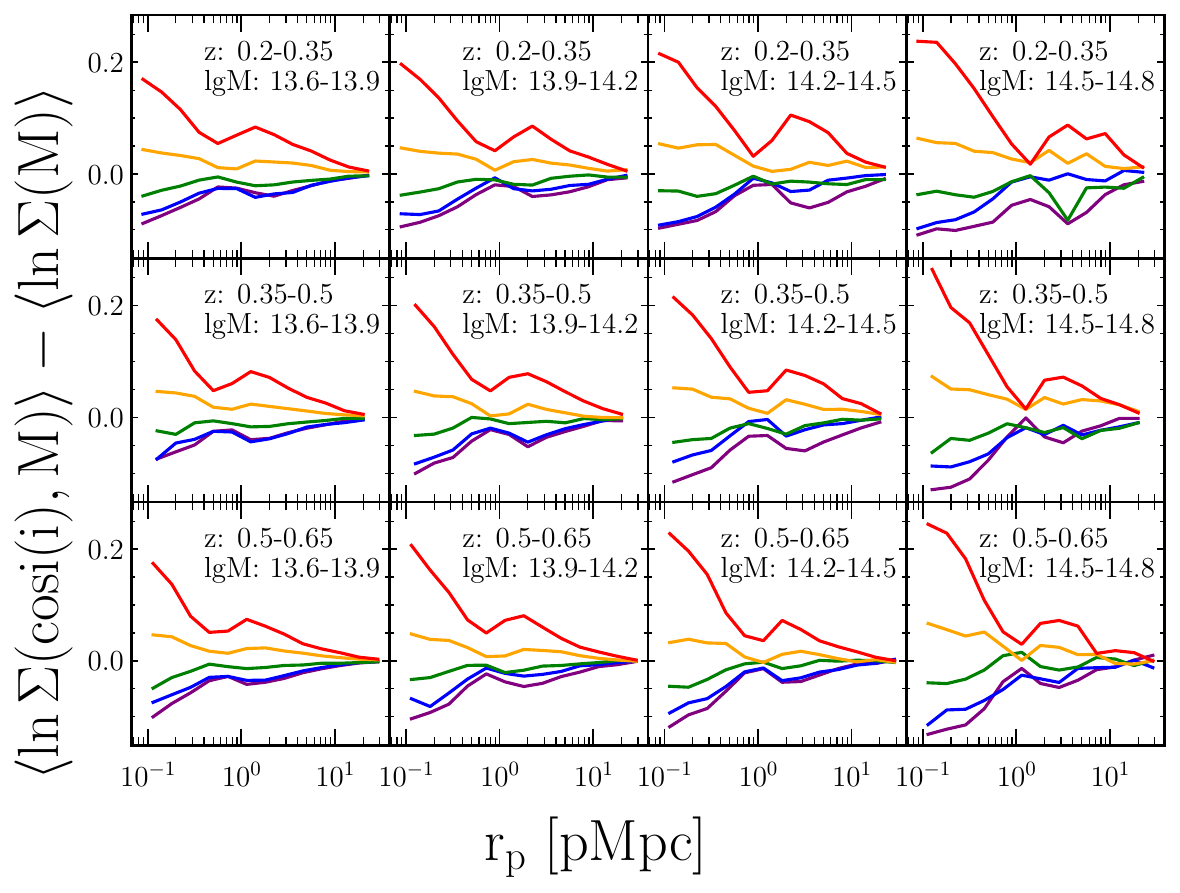}
\caption[]{Dependence of $\Sigma$ on halo orientation.  Each panel represents a {\em mass} and redshift bin, and the $\Sigma$ profiles are split into 5 cos(i) bins between 0 and 1.
}
\label{fig:Sigma_cosi}
\end{figure}
%%%%%%%%%%%%%%%%%%%

One of the possible sources of the selection bias is associated with preferentially selecting haloes with major axes parallel to the line of sight.  At a given halo mass, haloes with high $\cosi$ tend to have higher lensing signal than haloes of the same mass.  \cite{Osato18} show that the halo orientation can be associated with the enhanced lensing signal at both small and large scales (100 comoving $\hiMpc$ along the line of sight).  Using Buzzard simulations, \cite{ZZhang22} show that the \redmapper clusters indeed preferentially select haloes with major axes aligned along the line of sight.

Fig.~\ref{fig:cosi_pdf} presents the $\cosi$ distribution of our cluster sample, combining both Y1 and Y3 Buzzard realisations.  The orange histograms show the probability distribution of $\cosi$ for our richness-selected sample, and the blue histograms show that for the sample with the same mass and redshift distribution (based on the matching method in Appendix~\ref{app:diagnosis}).  The richness-selected sample includes more high-$\cosi$ haloes.

Fig.~\ref{fig:Sigma_cosi} shows how the $\Sigma$ profile depends on halo orientation.   Each panel corresponds to a redshift and mass bin.  For a given mass and redshift, we divide $\cosi$ into 5 bins of $\Delta(\cosi)=0.2$ (colour curves).  Let us focus on  $0.8 < \cosi < 1$ (red), i.e.~haloes with their major axes almost perfectly aligned with the line of sight. Their average $\Sigma$ profile is significantly boosted at small scales, has a dip at $\approx$ 1 pMpc, and has another peak at approximately 2--3 pMpc.  The selection bias we have seen in Fig.~\ref{fig:bias_DS}, however, has a different scale-dependence.    For high-richness clusters, the scale-dependence of selection bias is similar to what we see here.  However, for low-richness clusters, the selection bias is usually small at small radii and peaks at approximately 1 pMpc.  In addition, the 2--3 pMpc peak in Fig.~\ref{fig:Sigma_cosi} is at 5 -- 10\% level, which is lower than the amplitude in Fig.~\ref{fig:bias_DS}.  Therefore, while the orientation bias can account for part of the selection bias for the high-richness clusters, it cannot account for the selection bias for low-richness clusters because of the disagreement in the scale-dependence and the amplitude. It is possible that other orientation bias proxies \citep[e.g.][]{Dietrich14, Herbonnet21} could capture the selection bias more fully. For example, we have only calculated halo triaxiality at $\Rvir$; it is possible that a halo triaxiality proxy calculated at larger radii or a proxy of large-scale filaments connected to the halo could lead to a better model for selection bias.

%%%%%%%%%%%%%%%%%%%%
%%%%%%%%%%%%%%%%%%%%
%%%%%%%%%%%%%%%%%%%%
\section{Lensing selection bias associated with halo concentration}
\label{app:concentration}

%%%%%%%%%%%%%%%%%%%
\begin{figure}
\includegraphics[width=1\columnwidth]{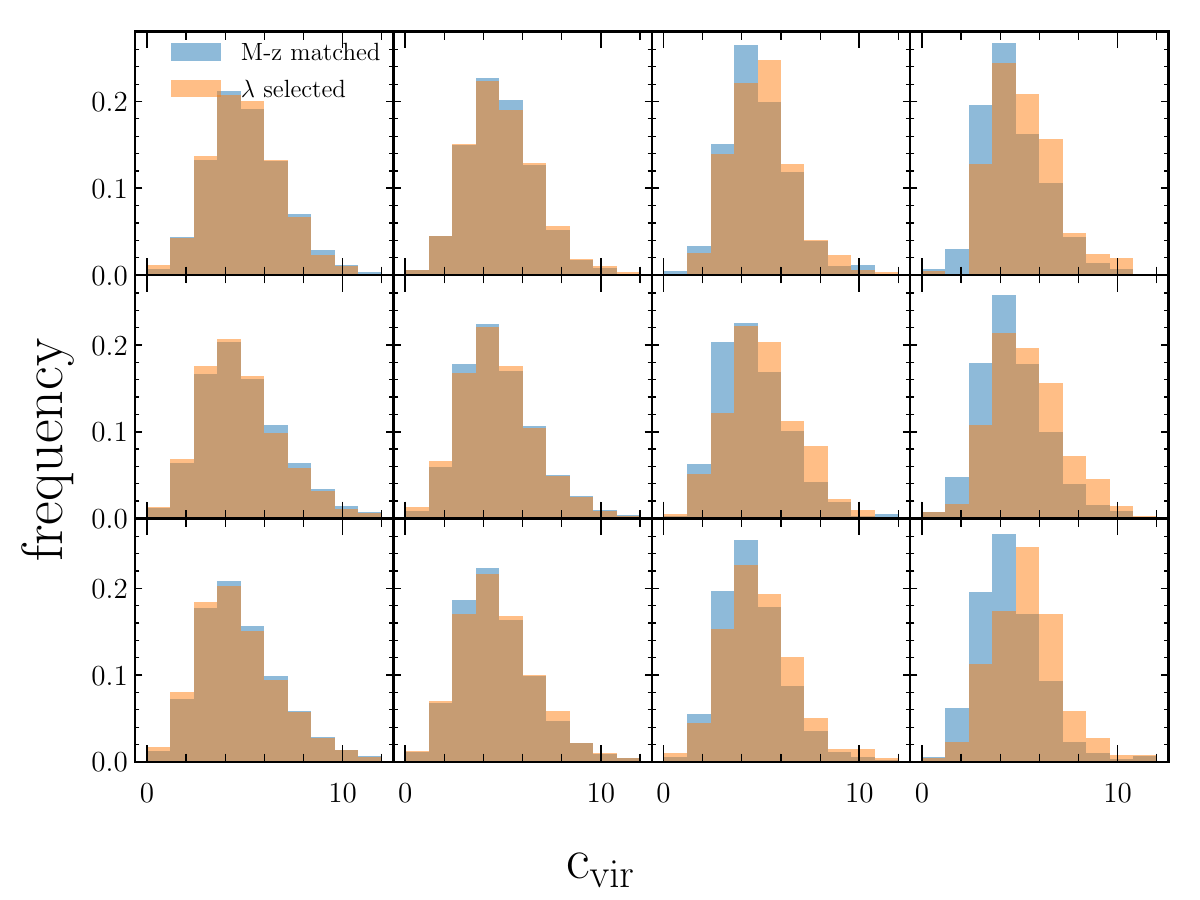}
\caption[]{Distribution for halo concentration for the richness-selected sample (orange) and a sample selected to match its mass and redshift distribution (blue). High-richness clusters show a stronger selection bias towards haloes with high concentration, while low-richness clusters do not show such a bias.}
\label{fig:cvir_pdf}
\end{figure}
%%%%%%%%%%%%%%%%%%%
\begin{figure}
\includegraphics[width=1\columnwidth]{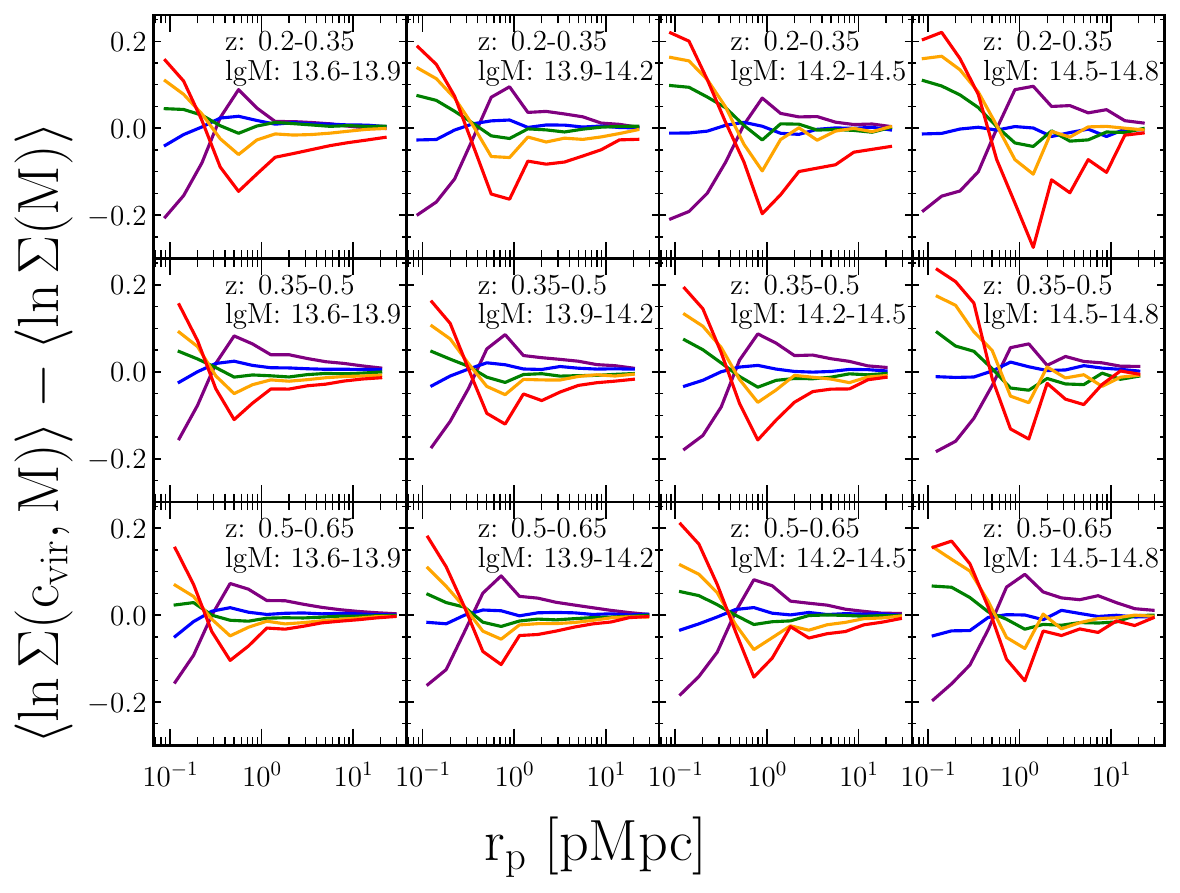}
\caption[]{Dependence of $\Sigma$ on halo concentration.  For a given mass and redshift, we split the haloes into five concentration quintiles.  Haloes in the highest concentration quintile have high $\Sigma$ for at $\rp \lesssim 0.5$ pMpc, which is compensated by a drop at $\rp \approx 1$ pMpc.}
\label{fig:Sigma_con}
%\vspace{-0.2cm}
\end{figure}
%%%%%%%%%%%%%%%%%%%
\begin{figure}
\includegraphics[width=1\columnwidth]{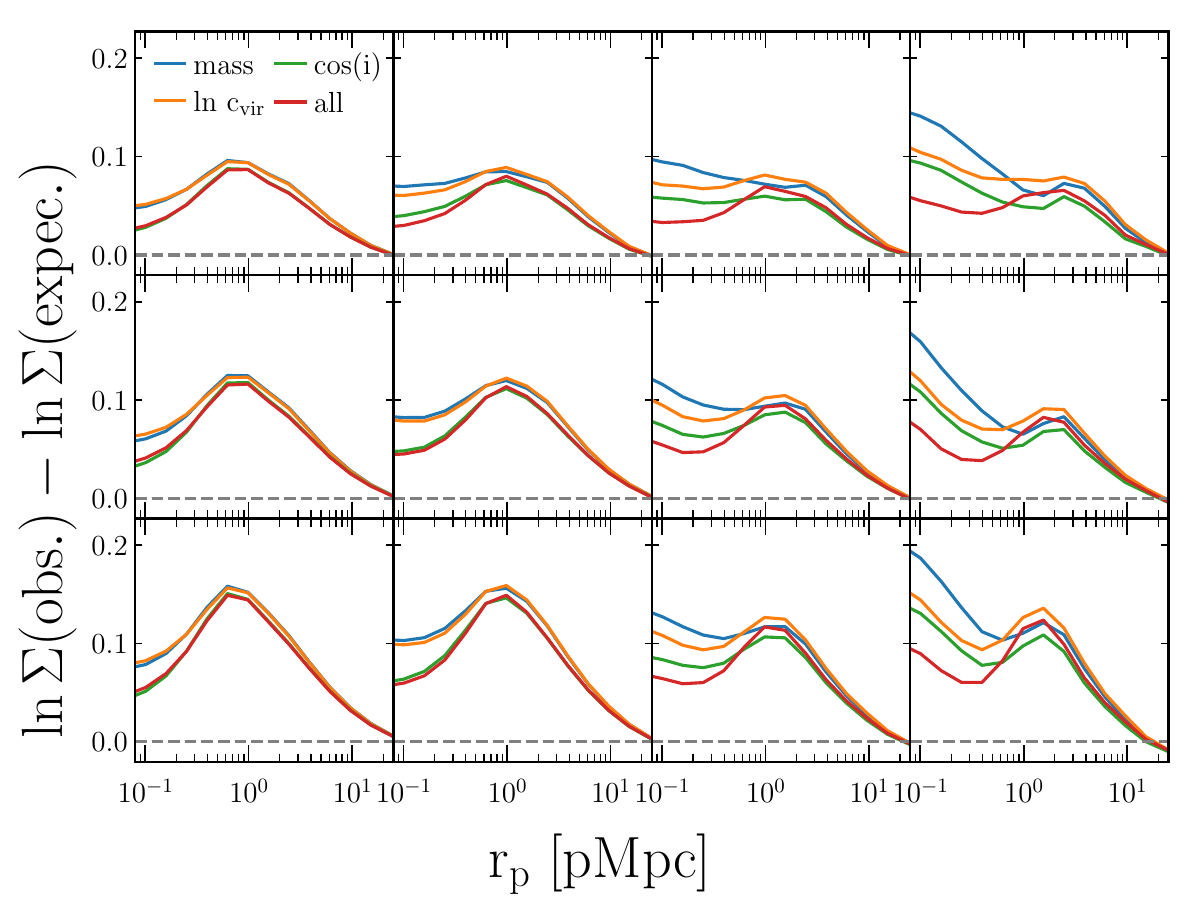}
\caption[]{Reduced lensing selection bias by accounting for the mass, $\cvir$, and $\cosi$ distribution.  The four curves correspond to taking into account mass (blue), mass and concentration (orange), mass and orientation (green), and all three properties (red).  The lensing selection bias is not eliminated.}
\label{fig:bias_Sigma_extra_prop}
\end{figure}
%%%%%%%%%%%%%%%%%%%%

In the main text, we have shown that high-richness clusters exhibit a selection bias in the 3D density profiles (Fig.~\ref{fig:bias_rho}).  In this appendix, we investigate the role of halo concentration.   Fig.~\ref{fig:cvir_pdf} shows the PDF of halo concentration, $\cvir = \Rvir / r_s$.  Here $r_s$ comes from the fitting of NFW profiles provided by the {\sc a} halo finder \citep{Behroozi13rs}.  As in Fig.~\ref{fig:cosi_pdf}, the orange histograms correspond to the richness-selected sample, and the blue histograms correspond to the mass-matched sample.   For high-richness clusters (the two right-hand columns), we see a clear preference for high-concentration clusters.  Such preference does not exist for low-richness clusters (the two left-hand columns).

Fig.~\ref{fig:Sigma_con} shows the dependence of $\Sigma$ on halo concentration.  For a given mass and redshift bin, we split the haloes into five concentration quintiles. For high concentration haloes, the small-scale behaviour is similar to high $\cosi$ haloes.  However, we can see that the scale-dependence for high-concentration haloes is very different from that of the selection bias.  Therefore, while the concentration could account for some of the high-richness selection bias, it plays a negligible role in the selection bias of low-richness clusters.

Fig.~\ref{fig:bias_Sigma_extra_prop} shows the reduction of selection bias when we match not only mass but also $\cosi$ and $\ln\cvir$ PDF when calculating the `expected' signal.  For this calculation, we generalise  equation~(\ref{eq:bias_corr}) to include $\ln\cvir$ and $\cosi$ in the linear regression.  The blue curves are our fiducial case that takes into account the mass, while the orange and green curves additionally take into account $\ln\cvir$ and $\cosi$, respectively.  The red curves take into account all three properties.  The selection bias is slightly reduced but is not eliminated.  This agrees with our reasoning that the lensing selection bias cannot be fully quantified by the biased selection of $\cosi$ and $\cvir$.

%%%%%%%%%%%%%%%%%%%%
%%%%%%%%%%%%%%%%%%%%
%%%%%%%%%%%%%%%%%%%%
\bsp	% typesetting comment
\label{lastpage}
\end{document}